# AI from Concrete to Abstract

## Demystifying Artificial Intelligence to the General Public

**Rubens Lacerda Queiroz[1] · Fábio Ferrentini Sampaio[2,3] · Cabral Lima[1,4] · Priscila Machado Vieira Lima[5,6]**

### Abstract

Artificial Intelligence (AI) has been adopted in a wide range of domains. This shows the imperative need to develop means to endow common people with a minimum understanding of what AI means. Combining visual programming and WiSARD weightless artificial neural networks, this article presents a new methodology, AI from concrete to abstract (AIcon2abs), to enable general people (including children) to achieve this goal. The main strategy adopted by is to promote a demystification of artificial intelligence via practical activities related to the development of *learning machines*, as well as through the observation of their learning process. Thus, it is possible to provide subjects with skills that contributes to making them insightful actors in debates and decisions involving the adoption of artificial intelligence mechanisms. Currently, existing approaches to the teaching of basic AI concepts through programming treat *machine intelligence* as an external element/module. After being trained, that external module is coupled to the main application being developed by the learners. In the methodology herein presented, both training and classification tasks are blocks that compose the main program, just as the other programming constructs. As a beneficial side effect of AIcon2abs, the difference between a program capable of learning from data and a conventional computer program becomes more evident. In addition, the simplicity of the WiSARD weightless artificial neural network model enables easy visualization and understanding of training and classification tasks internal realization.

✉ Rubens Lacerda Queiroz
rubensqueiroz@outlook.com

Fábio Ferrentini Sampaio
fabio.oxford@gmail.com

Cabral Lima
cabrallima@ufrj.br

Priscila Machado Vieira Lima
priscilamvl@cos.ufrj.br

1 PPGI
Federal University of Rio de Janeiro – UFRJ – Brazil

2 InovLabs – Portugal

3 Atlantica University – Portugal

4 Computer Science Department
Federal University of Rio de Janeiro – UFRJ – Brazil

5 PESC/COPPE
Federal University of Rio de Janeiro – UFRJ – Brazil

6 Tercio Pacitti Institute (NCE)
Federal University of Rio de Janeiro – UFRJ – Brazil

## 1. Introduction

In a recent article published by the Brazilian Academy of Sciences, Almeida (2018) states that:

> The preparation of a strategy for the advancement of artificial intelligence should start with some choices, such as: Which areas of application of artificial intelligence can generate the most economic growth and employment? How can artificial intelligence be applied to improve the quality of life of the Brazilian population? How to minimize the possible adverse effects of new technologies?

Almeida's questions are adherent to Yogeshwar (2018) position that "we need a culture where progress is the result of a reflection process of society, not the exclusive result of engineering and investors." He added, considering technological advances and their impacts on society: "most politicians do not understand what is happening, they are literally ignorant".

The perception of each person as an agent responsible for the future of technological development was also postulated by Medina (2004): "studies of the past […] reveal that human agency, not technological determinism, has governed the path of history and laid the groundwork for our current challenges".

An important question is: How would people, in general, and politicians, in particular, make choices, take decisions, set the course of our history *for the better* without understanding a minimum of the matter concerning their decisions, such as artificial intelligence (AI)? These decisions concern economic perspectives as well as moral and ethical aspects of this rapidly growing area.

An instance of the decision-making power of the general public concerning the future of AI is the recent public consultation launched by the Brazilian Government towards the development of a *Brazilian Artificial Intelligence Strategy*. The purpose of this consultation is "to submit to any citizen's contributions a set of questions that will guide a policy that enhances the benefits of AI in Brazil and the solution of concrete problems" (Caputo 2019). This initiative follows the example of consultations on AI carried out by other countries and by international organizations. In 2017, the European Parliament held a public consultation specifically on the future of robotics and artificial intelligence. "The public consultation included two separate questionnaires, adapted to their audience: one for the general public […] and one for specialists." (EUROPARL 2017)

People today are exposed to AI environments and they experience their potentials. However, the processes that allow this experience are not easily observable. Is the computer making inferences or the knowledge is showing what was explicitly *told* to it? How does it learn? One of the difficulties in endowing people with initial skills in AI lies in the complexity of techniques and concepts evolved in most AI systems (Sakulkueakulsuk et al. 2018).

Visual programming[1] has been used as a great tool for the learning of computer programming. Mixing visual programming and educational robotics lead to even more interesting results (Queiroz et al. 2019; De Luca et al. 2018; Chaudhary et al. 2016). Both approaches rely on the construction of knowledge from concrete references. This increases the potential audience of the target subject due to the mitigation of abstract reasoning. It is also possible to relate many aspects of AI to concrete references.

An essential part of AI that presents considerable understanding difficulty to the general public is machine learning. In its most basic form, machine learning consists of presenting examples of a certain class to an algorithm that will later classify new observations as belonging or not to that class. Current proposals for allowing the general public to experiment on machine learning and AI consist of two distinct phases. Block-based programming is included in that category. In the first phase, the learner uses an AI platform (e.g. IBM Watson[2]) to train a machine learning model. In the second one, he/she builds a block-based program that uses the model trained to classify new data. Classification results can then be used by decision structures in the program. This way, the intelligent part of the system is handled as a pre-existing abstract entity. However, as Richard Feynman said: "What I cannot create, I do not understand." (Caltech 1988), so it is important to find a simple way to explain artificial intelligence, especially machine learning, to the laymen.

To address Feynman's postulate, we propose to migrate learning/classification primitives to the same level as basic programming constructs. So, the beforehand *higher-level* tasks required for a machine to be able to learn from data are included as commands into the program being developed. The possibility of designing a programming language containing machine learning primitives was raised by Mitchell (2006). Such a language would allow the programmer to select a learning algorithm among the programing language primitives and the inputs to be learned and classified.

Combining both guidelines, we take machine intelligence from the world of abstractions and bring it closer to the universe of the manipulable. We also make easily visible the essential difference between a program that makes use of machine intelligence from one that just appropriate from the results from a third-party. Besides, the particular machine learning model adopted is possible to visualize through ludic activities of training and classification.

The possibility of building a basic understanding of artificial intelligence from concrete references presents an opportunity to bring some knowledge about this field to a wider variety of people, from distinct ages and backgrounds. To present AI from concrete to abstract methodology (AIcon2abs), this paper is organized as follows: Section 2 reviews related literature on explaining AI to the general public; Sections 3 and 4 describe the main concepts and technologies involved in the development of AIcon2abs methodology; Section 5 presents AIcon2abs itself; and finally, Section 6 discusses some final issues and points to future topics of research.

## 2. Related research

In (Queiroz et al. 2017) we singled out a work on AI teaching at the basic education level (Kandlhofer et al. 2016). That research used storytelling, unplugged computer activities[3], and educational robotics, to develop activities for teaching AI from kindergarten to high school. Those activities encompassed a lot of AI topics such as sorting and searching algorithms, graphs and data structures, intelligent agents, automata planning, and machine learning. Other references (Irgen-Gioro 2016; Blank et al. 2006; Greenwald et al. 2006; Parsons and Sklar 2004; Beer et al. 1999) focused mainly on teaching AI at undergraduate and graduate levels. Besides Gillian and Paradiso (2014) and Marshall (2004) also proposed the development of machine learning tools dedicated to aid professionals of other areas, such as artists. The main strategies adopted were based on educational robotics and agent-oriented program.

Kandlhofer et al. (2016) identified that teaching basic concepts and techniques of AI at the school level was quite rare. This still seems valid once the amount of scientific research in this field remains somewhat unexpressive.

[1] "When a programming language's (semantically significant) syntax includes visual expressions, the programming language is a visual programming language." (Burnett 2002, pp. 77)

[2] https://www.ibm.com/watson

[3] Activities used to teach computer science concepts without a computer (Bell et al. 2009).

Due to the scarcity of new scientific work in this area, our search also covered tools developed outside the scientific environment. Subsection 2.1 summarize the main references found through *Periódicos Capes*[4], while Subsection 2.2 presents the most significant works found by the internet search using the words: ("artificial intelligence" OR "computational intelligence" OR "machine learning") AND (fundamentals OR "basic concepts" OR beginners OR dummies OR children OR "k-12" OR "primary school" OR laymen OR "lay people").

## 2.1. Scientific databases search

Hitron et al. (2018) carried out a research with children (10-12 years old) to observe if they were able to identify two basic machine learning concepts: data labeling and data evaluation. Their experiment used a tennis-like movement recognition device. The results showed that the subjects were able to understand the desired concepts and to identify their application in real-life devices, such as autonomous cars and smart speakers.

On another work, Hitron et al. (2019) proposed opening machine learning black boxes to children (10-13 years old). Education activities focused on classification tasks, considered by the authors as being less complex and more common in real-world applications than other kinds of machine learning problems. In their research, the process of learning gestures was performed in 2 steps: (i) the learners fed the machine learning model with positive and negative examples of the class of gestures being learned; (ii) the students evaluated the trained model accuracy on the recognition of new examples. The application interface had one button to change from the training phase to the recognition one. The user could toggle from one view to the other to retrain the model and make further observations of the new results. Feature extraction and model selection were not included in that process due to the complexity.

Sakulkueakulsuk et al. (2018) proposed to introduce AI to middle-school students in Thailand, which has 49 percent of labor employed in the agriculture sector. The approach integrated machine learning, gamification, and social context in STEM[5] education. Using the platform Rapidminer[6], the students had to train different models with features such as texture and color of mango, and observe which of the models would better predict, for example, the flavor of the mango based on its external features.

Druga (2018) explored how children (7-14 years old) improved their understanding of AI concepts and how they changed their perception of smart systems using Cognimates.me[7]. The platform uses Clarifai[8] and Uclassify[9] for people to train machine-learning datasets with images or text. Once trained, these models can be integrated into programs developed in Scratch[10]. Cognimates.me also provides extensions for Scratch that allow people to interact with smart systems, such as Amazon Alexa, and with pre-trained machine learning models to recognize, for example, colors or whether a text is positive or negative.

## 2.2. Google search

Deep learning is a branch of machine learning developed for solving complex tasks such as image classification, character and speech recognition, object identification, machine vision, and others. Deep learning algorithms integrate several mathematical, logical, and computational methods and require a large amount of data from which to learn (Vasuki and Govindaraju 2017, Graupe 2019, Gil et al. 2020). Due to their remarkable performance, deep learning approaches have widened the use of AI solutions, giving birth to an increasing number of sites and courses that aim to teach AI to children or laypeople. Some instances of that are: *AI in Schools*[11], *Teens in AI*[12], and *Machine Learning for Kids*[13].

NVIDIA[14] designed the course AI in Schools[11] for helping teachers to demystify the topic of artificial intelligence to their students. Among other activities, the learner can use a free access web platform to train a deep machine learning model to classify images.

Teens in AI[12] is a methodology designed to expose young people (12-18 years old) to AI technologies for social good. Some of the topics covered by the project are AI, machine learning, and data science. The approach combines a set of activities such as hackathons, boot camps, and accelerators with expert mentoring.

Machine Learning for Kids[13] presents a web platform to work AI concepts with children. By using IBM Watson[2], children can train machine learning models with text, image, sounds and numbers. The learners are able to use the trained models to develop smart systems using Scratch[10].

[4] *Periódicos Capes* is a Brazilian search engine that searches for papers in many international scientific databases such as SprigerLink, ACM Digital Library, IEEE explore, Scopus, ScienceDirect, SciELO and Google Scholar. http://www.periodicos.capes.gov.br
[5] STEM stands for Science, Technology, Engineering and Math.
[6] https://rapidminer.com/
[7] http://cognimates.me/home/
[8] https://www.clarifai.com/
[9] https://www.uclassify.com/
[10] https://scratch.mit.edu/
[11] http://aiinschools.com/
[12] http://teensinai.com
[13] https://machinelearningforkids.co.uk/
[14] https://www.nvidia.com/

In addition to the above examples, there are other sites, magazine articles, YouTube videos, and blogs dedicated to artificial intelligence and the general public. The AAAI and Computer Science Teachers Association (CSTA) (AAAI 2018), in order to demystify AI to the general public, organized a group to develop national guidelines for teaching AI to K-12 students. This exemplifies the growing interest in AI in the whole society.

## 2.3. State-of-the-art summary

In the works presented in Subsections 2.1 and 2.2, a quite common strategy employed was to use an external machine learning model for training and to evaluate its quality via observing the classifications performed. Two of the approaches, Cognimates.me (Druga 2018) and Machine Learning for Kids[13], include the building of block-based programs using the previously trained models. These approaches are the closest to AIcon2abs methodology, and Table 1 shows features contemplated or not by them.

Table 1: Cognimates.me (CN) and Machine Learning for Kids (ML) features

| Features | CN | ML |
|---|---|---|
| Model training is fully performed as part of block-based programming | | |
| Image learning and recognition requires internet | ✓ | ✓ |
| Allows consistent generalizations from a single example | | |
| Online learning process (allows for interleaving between training and classification tasks) | | |
| Uses a third-party AI platform for image training and classification | ✓ | ✓ |
| Uses a simple learning and recognition model that can be replicated in its original form by students in ludic activities | | |
| The learning model adopted allows the visualization, as a concrete image, of what was learned | | |
| The block environment has a downsized set of blocks | | |
| The Block environment has individual personalized icons for each block | | |
| Capable of controlling Raspberry Pi [15] and Arduino[16] GPIOs[17] ? | ✓ | ✓ |
| Allows image learning and recognition | ✓ | ✓ |
| Allows text learning and recognition | ✓ | ✓ |
| Allows sound learning and recognition | | ✓ |
| Allows controlling sprites on computer screen | ✓ | ✓ |
| Text to speech without internet | | |
| Presents theories about cognitive maturity that support the approach in question | | |

Those approaches require an internet connection and do not include training the model as part of the programming environment. Besides, they do not present a study about the impact of the learner's cognitive maturity on the desired learning. AIcon2abs will be described in Section 5 and summarized in Table 2 of Section 6.

## 3. Concepts involved

Design Science Research (Hevner and Chatterjee 2010; Wieringa 2014; Pimentel et al. 2019) suggests that the construction of a computational artifact, developed to solve a given problem, needs to be driven by theoretical conjectures. In turn, the artifact in question is used to check whether these conjectures hold (Figure 1).

The use of the artifact allows evaluating whether the conjectures hold

Artifact

Implementation of an artifact based on theoretical conjectures

Theoretical Conjectures

Conjectures about human behavior that must be taken into account when designing an artifact

Theoretical conjectures guide the design of the artifact

Figure 1: Correlation between conjectures and artifact (Pimentel et al. 2019)

Theoretical conjectures are assumptions about human, organizational, or social behavior. They are based on theories, models, and constructs from various areas. "The theoretical conjectures are not usually found in Computing, but in other areas, such as Education, Psychology, Communication, Cyberculture, etc." (Pimentel et al. 2019, pp. 25)

AIcon2abs is based on DuinoBlocks4Kids (DB4K) (Queiroz et al. 2016, 2019), a kit for teaching computer programming to children through educational robotics. DB4K kit was developed from the following central theoretical conjecture: *The learning of some computer science abstract concepts can be facilitated through practices based on aspects of these concepts that can be easily observed in the concrete world*. The tests carried out with DB4K pointed this conjecture to be valid.

Subsections 3.1 and 3.2 presents education and psychology theoretical fundamentals that served as the base for both AIcon2abs and DB4K. Subsections 3.3 and 3.4 introduce the concepts necessary to endow AIcon2abs with machine learning primitives.

[15] Raspberry Pi is a low-cost single-board computer provided with a set of General Purpose Input/Output pins where one can connect robotics devices. https://www.raspberrypi.org/

[16] Arduino is a low-cost, open-source electronic prototyping platform that is simple to use for any student, including children. http://www.arduino.cc/

[17] In the context of educational robotics, GPIO (general-purpose input/output) is roughly a standard electronic interface (commonly known as input/output pin) used to connect sensors and actuators, such as light sensors and LEDs, to an electronic control board.

### 3.1. Constructivism, Constructionism, and knowledge construction

Piaget's Constructivist theory (1950) shows that the increase of knowledge is built from the interaction of the individual with the physical environment. Papert's Constructionist theory (1993a) added to the Piagetian Theory the idea that the construction of knowledge takes place more effectively when the learner consciously engages in the construction of something tangible. Papert perceived the computer as a tool able to expand the possibilities of creation and, consequently, learning. The idea is that the computer allows the development of projects with a higher degree of complexity than those the child would be able to build using only the *physical world* (Papert, 1993a).

Within this context, Papert created Logo, a software that allows users, through lines of code, to guide a *turtle*. The turtle is a cybernetic animal that can be either a virtual object (on a computer screen) or a manipulable physical object (Papert 1993a). This turtle leaves a trail (a drawn line) while it walks, allowing the user to have immediate feedback on the commands given by him/her to the computer. The construction of knowledge is achieved through the user's reflection on the concrete results of the commands given by him/her to the computer from the observation of the graphic elements constructed by the turtle's movements.

These same Constructivist and Constructionist principles can be adopted to develop a baseline understanding of AI in the general public. Using block-based programing, people can build *learning machines* and interact with the machines they have built. As a result, an understanding of the machine learning process can be developed from the observation of the behavior presented by the systems that the learners themselves have developed. Besides, using WiSARD (Wilkie, Stonham, and Aleksander's Recognition Device) paradigm, learners can also visually observe and replicate, through ludic activities, the internal process performed by the machine when learning. Each of these activities can be followed by discussions about AI fundamentals, reflections of its application in society, the responsibility of each citizen about the future of artificial intelligence, among others.

### 3.2. Cognitive maturity and the power of abstraction

Jean Piaget distinguishes four general stages in cognitive development, namely: sensorimotor (0-2 years old), preoperational (2-7 years old), concrete-operational (7-11/12 years old), and formal-operational (11/12 years old and beyond). These age groups are just a reference. All individuals progress through these stages in this exact order. However, according to Piaget, there is a wide variation in the ages at which a person enters or emerges from each stage. Cultural and environmental factors may either quicken or delay the speed of an individual's intellectual growth (Shaffer 1996).

The concrete-operational stage starts around the age of seven. In this stage the subject develops the ability to "mentally representing what has already been absorbed on the level of action" (Piaget and Inhelder 1969, pp. 94). In other words, "physical actions begin to be *internalized* as mental actions or *operations*" (Beard 2006, pp. 76).

For Piaget, concrete, in the sense of concrete-operational stage, means that the mental operations are built from some aspect of external reality: (i) physically present, (ii) already experienced or (iii) mentally represented. That is, in the concrete-operational stage, knowledge is built from what is observable, not from a definition. The individual needs to compare what is being learned with what is already known or is being physically perceived. Thus, hands-on activities should be used to present scientific concepts to people in the concrete-operational stage, since they are capable of reasoning mostly about the concrete and the manipulable (Brown et al. 1996; Pedrozo 2014).

Therefore, empirical is the dominant kind of abstraction in this stage. It consists of the construction of reasoning from the abstraction of objects belonging to the universe of the person in question. Empirical abstraction emerges from the observation of physical objects or from material aspects of actions, such as movements (Piaget et al. 1980). For individuals in the concrete-operational stage, the construction of empirical abstraction is, in general, a routine task. In hypothetical-deductive thinking it would be necessary to construct abstractions from hypotheses (reflexive abstraction). This kind of abstraction tends not to appear during concrete-operational stage (Lister 2011).

As mentioned before, a sort of factors may cause subjects of the same age to be in distinct cognitive maturities levels (Shaffer 1996). Besides that, a same individual may have characteristic cognitive traits of different developmental stages concurrently. Even adult individuals may not have the capacity for abstraction fully developed (Kramer 2007). Tests conducted on adult populations indicated that only about 30% of adults achieve formal operational skills, which includes the ability to perform reflexive abstraction. Most adults remain in a transitory stage between concrete and formal operations (Kuhn 1977).

Furthermore, according to the neo-Piagetian theory, regardless of their age, people's power of abstraction on a specific domain increases along with that person's experience concerning that domain. "Thus, a person who is a novice in one domain (e.g. chess) will exhibit less abstract forms of reasoning than that same person will exhibit in a domain where he is expert (e.g. calculus)." (Lister 2011, pp. 10)

> […] when facing the need to cope meaningfully with concepts that are too abstract for them, CS [(Computer Science)] students tend to reduce the level of abstraction in order to make these abstract concepts meaningful and mentally accessible […] by dealing with specific examples instead of with a whole set defined in general terms (Hazzan 2008, pp. 40).

Consequently, the adoption of approaches that require less abstraction capacity from the subject to understand AI concepts becomes appealing for people in general.

## 3.3. Perception of intelligence

Discussing the answer to the question *Can machines think?* Turing (1950) presents a game called *The Imitation Game*. In short, the original game works as follows: three people, one man (A), one woman (B) and one interrogator (C) participate in the game. Separate from the couple, now baptized as X and Y, the interrogator (C) can ask them any question. X and Y must answer the questions through typed papers to prevent the tone of the voice or the form of the writing from helping to identify *who is who*. The Interrogator's goal is to find out if X is the man (A) and Y is the woman (B) or vice versa. During this process, A must try to induce C to lose the game by pretending to be B, and B must help C by attempting to show that she is B.

Turing proposes a modification in the game. In the new version, a machine would take the place of A and try to deceive the interrogator by pretending to be human. Meanwhile, B (now male or female) would try to help the interrogator to distinguish who is the machine and who is the human *behind the wall*.

> What will happen when a machine takes the part of A in this game? Will the interrogator decide wrongly as often when the game is played like this as he does when the game is played between a man and a woman? These questions replace our original, 'Can machines think?' (Turing 1950, pp. 434).

Looking for answers to these questions, Turing presents an in-depth discussion on themes surrounding the possibility of the existence of intelligent machines, including aspects of different areas such as philosophy, mathematics, biology, religion, and psychology. But somehow, in current context of artificial intelligence, the ideas and concepts brought by Turing were reduced to a test named *The Turing Test*. Based on the Imitation Game, the test aims to determine if a machine is intelligent or not. However, the value of the Turing Test as a benchmark for artificial intelligence, as the goal of AI, is widely criticized by AI researchers and experts. According to Stuart Russell: "Almost nobody in AI is working on passing the Turing Test, except maybe as a hobby," and people working on passing the test would not be described as mainstream AI researchers (Prado 2015). Marvin Minsky called the Loebner Prize, the world's oldest Turing Test competition, "obnoxious and stupid" (Dormehl 2017). For Marcus (2014): "In terms of practical significance for artificial intelligence, […] passing the Turing Test means little."

The value of the Imitation Game is not in its use as a test to verify if a machine can think like a human being but in understanding AI. Through the proposal of the Imitation Game, Turing gives us the idea that what matters in our understanding of whether a machine is *thinking* is if it behaves intellectually as a human would behave. It would not matter what the machine is doing to act that way, neither if the term *thinking* is the most appropriate to the internal processes the machine is performing to produce the observed outputs. The important thing is that we have the perception that the machine is thinking. In other words, it is all a matter of intelligence perception from behavior, as emphasized by Stuart Russel:

> [The Imitation Game] was designed as a thought experiment to explain to people […] that the possibility of intelligent machines did not depend on achieving consciousness […], an argument about the importance of behavior in judging intelligence […] (Prado 2015).

This approach supports the idea of developing a baseline understanding of AI from the observation of the differences between the behavior performed by *intelligent machines* and those presented by conventional computer programs. In other words, we propose to use the Turing Test not just as a tool for evaluating the performance of intelligent systems, but also as a tool for the comprehension of artificial intelligence through comparing the behavior expected from a human from the behavior of an artificial system.

## 3.4. Intelligent agents and the learning process

Russell and Norvig (2010, pp. 35) present a concise definition of agent: "An agent is anything that can be viewed as perceiving its environment through sensors and acting upon that environment through actuators." This concept enables a clear establishment of a relationship with elements of everyday life. The human being has ears, eyes, and skin as some of his/her sensors and mouth, arms, and other parts of his/her body as actuators. A robot may have cameras and ultrasonic devices as sensors and motors as actuators. A computer program uses the keyboard and files as sensors, and the screen, speaker, and printer as actuators (Russell and Norvig 2010).

The agent concept is used as a tool to design/analyze intelligent systems. This analysis can be conducted by observing the actions performed by the agent actuators as a response to the inputs received by its sensors. An intelligent agent learns from what it perceives from the environment, possibly changing its behavior as its knowledge about the world is improved (Russell and Norvig 2010). The agent-based approach can be used to reduce the power of abstraction needed to analyze the difference between an intelligent system and a conventional computer program because it is based on the observation of external reality aspects, either physically present or already experienced.

Mitchell (1997, pp. 2) proposes an interesting definition for the learning of an artificial intelligent agent:

> A computer program is said to learn from experience *E* with respect to some class of tasks *T* and performance measure *P*, if its performance at tasks in *T*, as measured by *P*, improves with experience *E*.

This definition can be applied to a variety of learning tasks, for example, the handwriting recognition learning problem, in which we can identify (Mitchell 1997, pp. 3):

- Task *T*: recognizing and classifying handwritten words within images
- Performance measure *P*: percent of words correctly classified
- Training experience *E*: a database of handwritten words with given classifications

By applying this concept, the machine learning process is perceived through the observation of the agent's behavior. The difference between an intelligent system and a conventional computer program can be analyzed based on observable aspects of *intelligence*.

## 4. Technologies involved

In addition to being guided by theoretical conjectures, the conception of a computational artifact is also based on techniques, design principles, and related artifacts (Pimentel et al. 2019). In this section we present the two main technologies that support AIcon2abs methodology: DuinoBlocks4Kids (Queiroz et al. 2016, 2019) and WiSARD (Aleksander et al. 1984).

DB4K is a didactic kit for teaching computer programming via educational robotics based on the use of free technology, low-cost components, and recyclable materials. The kit consists of a block-based visual programming environment, a set of educational robotics materials, and a collection of activities. Its use is focused on exercising some *Computational Thinking skills* (Wing 2006) in elementary school children (7 years old onwards).

WiSARD is a lightweight Weightless Artificial Neural Network (WANN) model. It has very pictorial and simple learning and recognition primitives. The model was originally designed for image recognition but currently finds application in various domains (Lusquino Filho et al. 2020; Santiago et al. 2020; Carneiro et al. 2017; De Gregorio and Giordano 2017; Cardoso et al. 2016; Carneiro et al. 2015; De Gregorio et al. 2012). We will present in this section the original WiSARD model as well as one with a more efficient tie-break mechanism and the ability to produce prototypes of what has been learned.

### 4.1. DuinoBlocks4Kids

DuinoBlocks4Kids (Queiroz et al. 2016, 2019) was built on constructivist assumptions, especially the understanding that children aged 7 to 12 are in the concrete-operational stage (Shaffer 1996). In this stage, as described in Subsection 3.2, the subject resorts to concrete objects present or already experienced to perform mental operations. Another theory that guided the development of the kit is Papert's Constructionism (1993a), which combines Piaget's Constructivist theory (1950) with the use of computers in education. Figure 2 shows the concepts and technologies involved in the development of DB4K.

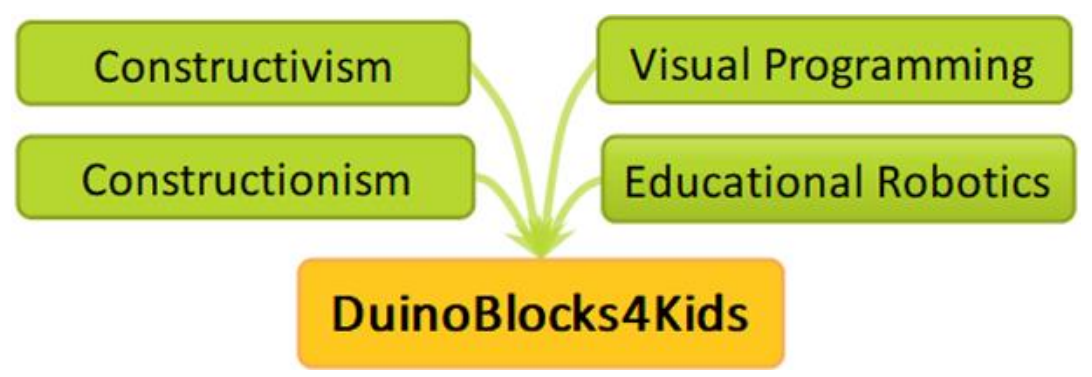

Figure 2: Concepts and technologies involved in the development of DB4K

DB4K visual block-based programming environment (Figure 3) is a programming environment for Arduino[16] boards developed using Blockly[18] and Ardublockly[19]. The use of block-based programming was popularized by Scratch[10], a project of the Lifelong Kindergarten Group from MIT Media Lab. Through the DB4K available blocks, it is possible to control de most common robotics devices used in educational robotics classes with Arduino. There are also blocks responsible for the program flow control, such as Repetition and Decision structures.

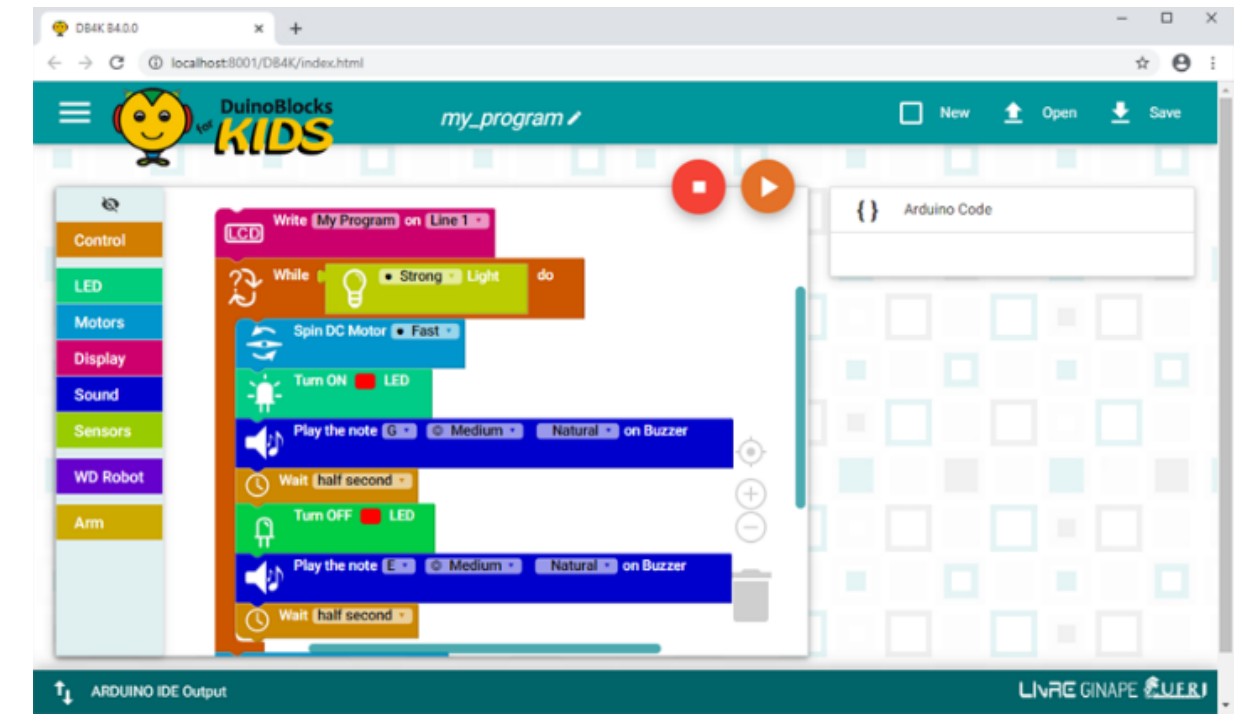

Figure 3: DuinoBlocks4Kids visual programming environment[20]

[18] https://developers.google.com/blockly/
[19] https://ardublockly.embeddedlog.com/index.htmlt

[20] This is an adapted image. The original DB4K interface has texts in Portuguese. http://ginape.nce.ufrj.br/LIVRE/paginas/db4k/db4k.html

DB4K blocks were designed to aid their use by children, in contrast with the blocks commonly found in visual environments for Arduino boards programming. This is achieved through a combination of text and icons. Each block makes explicit the device it controls and the result to be observed when the block is executed in the program. Hardware-related details, such as pinouts and voltage level values, are suppressed. To light up an LED, block-based programming environments for Arduino usually use the block: *Set Digital pin number < n > to < Hight/Low >*. DB4K uses the block *Turn on < color > LED*, where the LED is identified by its color (Figure 4). That is, the device controlled by the block is an LED, and the expected result of using that block to control this device is that it lights up.

The parameters used in the blocks were also simplified. In the case of the block *Spin DC Motor < velocity >*, instead of having to enter a numeric value between 0 and 255 to be applied to a given digital pin, the child chooses one of three predefined speeds within his/her universe of understanding (Figure 5).

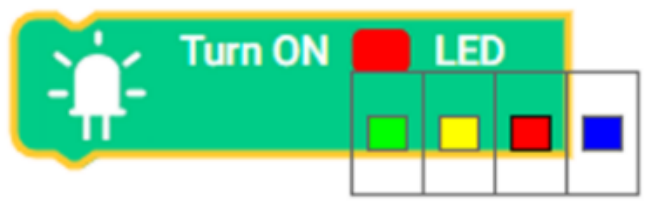

Figure 4: *Turn ON < color > LED* block [20]

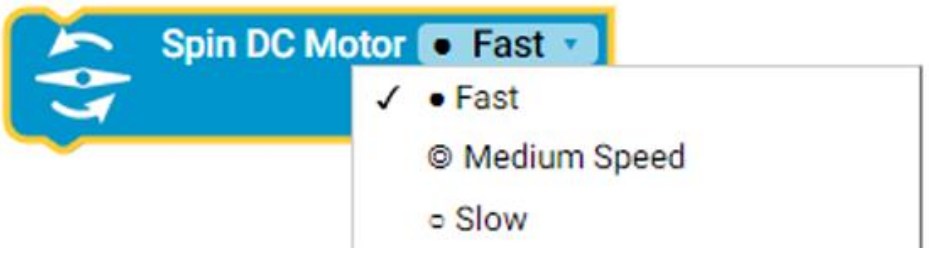

Figure 5: *Spin DC Motor < velocity >* block [20]

The kit is also composed of a set of educational robotics materials. One of these materials is a plastic box, named *The Little Magic Box* (Figure 6) which has all devices programmable by DB4k already connected to an Arduino Board. Some PET bottle robots were also developed, such as the Robotic Bat and the Robotic Fish (Figure 7). These robots are employed in activities with narratives to contextualize the use of the robotics devices previously dealt with in class with The Little Magic Box. The use of these materials, along with the block-based programming environment, enables children to perform the debugging process. Through this process, the learners can, on their own, discover, verify, and correct possible errors in the logic of the developed programs. A particularly important task as it enables children to have more autonomy in their learning.

The DB4K kit was used in a study carried out with seven children (five boys and two girls) living in a low-income community in Rio de Janeiro, Brazil. The students had no previous computer programming experience and were enrolled in public schools. Four of them were from the 4th grade and three from the 3rd grade. The workshop was composed of 14 meetings of 90 minutes each. The results of the workshop pointed out the feasibility of working the following skills of computational thinking with children from seven years onwards by using the DB4K kit: Ability to perform abstractions (more specifically empirical abstraction), understanding control flows, debugging and systematic error detection, iterative thinking, use of conditional logic, and structured problem decomposition.

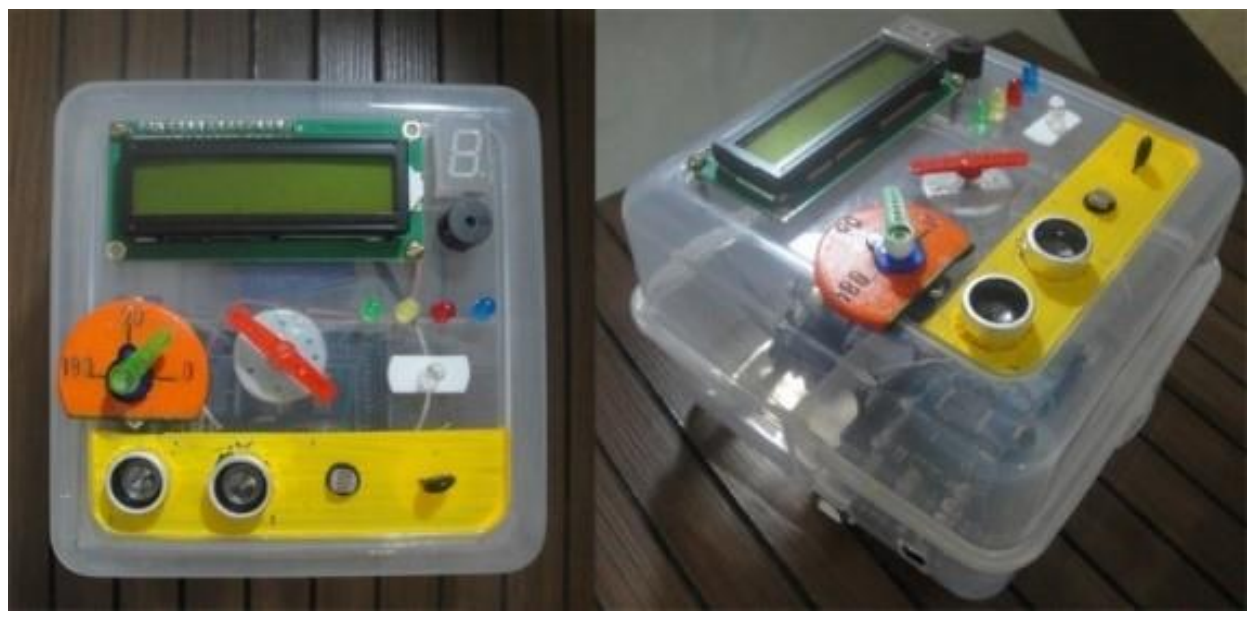

Figure 6: Little Magic Box

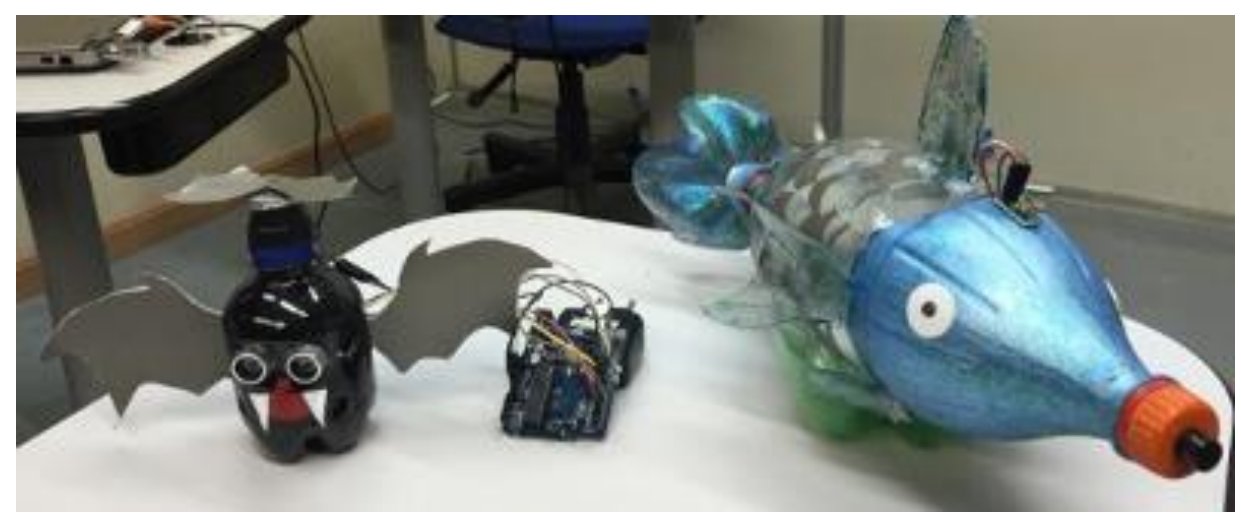

Figure 7: PET bottle Robotic Bat and Robotic Fish

DB4K presents itself as an appropriate choice from which to develop a block-based programming environment for demystifying AI to the general public. It was built respecting the abstraction capacity and other cognitive abilities of subjects in the concrete-operational stage. As mentioned in Subsection 3.2, reducing the level of abstraction needed to understand contents related to domains with which you are not familiar makes this understanding more accessible not only to children but also to a significant portion of the adult audience.

## 4.2. Original WiSARD

In this section, we will present concepts and mechanisms of the original WiSARD (Aleksander et al. 2009) that enable an easy visualization and reproduction of its training and classification processes through ludic activities. Additional details of WiSARD artificial weightless neural networks, as well as the differentiation between weightless and conventional (weighted) artificial neural networks, can be found in Aleksander et al (1984), Aleksander et al. (2009), França et al. (2014) and Lima Filho et al. (2020). The motivation for using WiSARD as the machine learning engine for AI demystification to the general public came from research on the use of this model in image recognition performed around the paper *Playing with Robots Using Your Brain* (Queiroz et al. 2018).

WiSARD is a machine learning device that has its training and classification process based on writing and reading data from computer memoires of type RAM (Random-access Memory) (Dennard 1968; Mano 1979). It is possible to understand the computer memory as a set of little boxes where each one can be used to store only a 0 or 1 value, called bits. The entries for WISARD, i.e., what it will learn, must be in binary format as in the examples of black-and-white images that will be presented in this section. As mentioned earlier, WiSARD was originally developed for the image recognition task, the same type of task adopted by AIcon2abs methodology. However, different techniques can be used to binarize non-binary inputs. These inputs are then used to train/test WiSARD machines in different domains. WiSARD has two operations modes/phases: Training and Classification. Inputs in the Training phase means that some information will be written in the WiSARD RAMs, whereas input data in the Classification phase means that information will be read from RAMs.

WiSARD is essentially composed of the following elements:

- *Retina:* A vector of binary pixels used as the input to WiSARD. The vector is usually arranged as a bidimensional matrix such as it is displayed in figure 8.
- *Neurons:* RAM memories.
- *Discriminators:* A set of neurons that hold knowledge about a class of data. For example, a letter, an animal species, the meaning of a sound, and so on.
- *Mapping:* It is a correspondence between the pixels of the retina and the bits that compose de address of each RAM.
- *Adders:* Element employed to indicate the degree of similarity of an observation that appears as inputs in the classification phase from a certain class WiSARD has already learned.

As a first example, we will use the learning of the letter E pattern. It will be represented by a black and white image of 3 by 5 pixels[21] (Figure 8). A black pixel corresponds to the number 1, and a white pixel corresponds to the number 0. In this example, we have a *Retina* of 15 pixels that can be black (equal to 1) or white (equal to 0).

The first step WiSARD needs to perform is to divide the Retina into sets of pixels sequences named *Tuples*[22]. WiSARD uses these Tuples to *Map* the pixels of the Retina to bits that address a certain position to be accessed in the *Neurons* (Figure 11). It is desirable that the number of pixels that will compose each Tuple is a divisor of the total number of Retinal pixels. Otherwise, we will have to deal differently with the remaining. In this example, we will use Tuples of 3 pixels. To simplify the visualization we will identify each of the retina's pixels by their coordinates, as in a Battleship game (Figure 9).

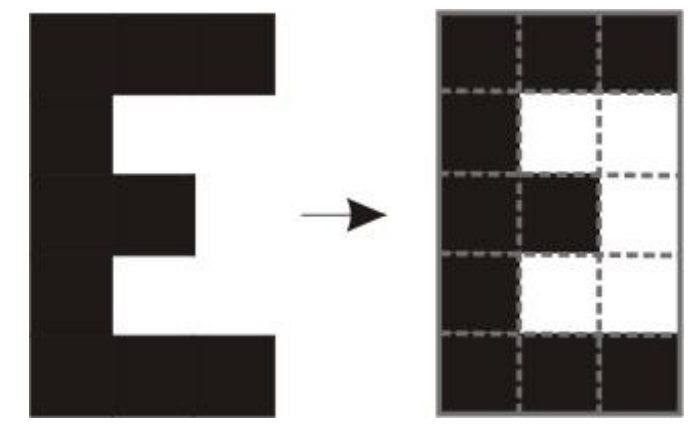

Figure 8: Letter E in 3x5 pixels format

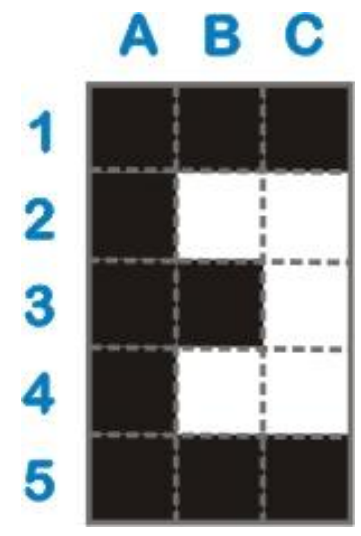

Figure 9: Retinal pixels coordinates

The sequence of pixels that will be part of each Tuple is chosen at random[23]. For the example presented here, WiSARD will use the following Tuples: (A4, B2, C1), (A1, C4, A5), (C3, A2, B4), (B3, C5, A3) and (C2, B1, B5). Each position of a Neuron is identified by an address. The number of elements in each Tuple determines the number of digits that will compose these addresses. The number of positions in a Neuron will be equal to 2 raised to the power of the number of digits of each address. In this example, each address is composed of 3 digits. So, 8 ($2^3$) positions can be addressed, corresponding to the addresses: 000, 001, 010, 011, 100, 101, 110, 111(Figure 10).

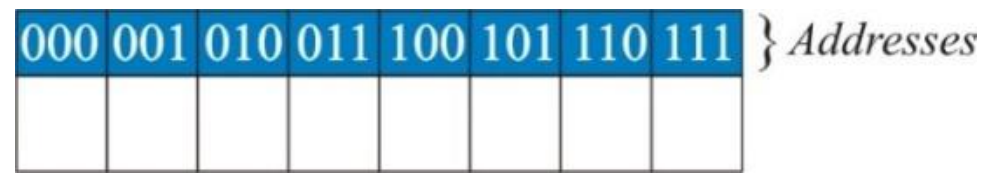

Figure 10: A WiSARD neuron with 8 positions

Each Tuple in the mapping indicates an address on one of the Discriminator's Neurons. A *Discriminator* is formed by the set of Neurons that will hold information about a specific class of data to be learned by WiSARD (which in this example are letters). In this case, as we have 5 Tuples, we will have Discriminators formed by 5 Neurons (Figure 11). With that, we have everything we need to teach a letter to our WiSARD: A Retina, a Mapping, and a Discriminator formed by a set of Neurons.

The *Training Phase* consists of writing a number 1 in the memory positions of each Neuron indicated by the addresses formed by the Tuples. At the beginning of Training, all positions of all Neurons are filled with the value 0. So, based on the five Tuples presented above, to learn the letter E example presented in Figure 11, WiSARD will write a number 1 in the position indicated by the pixels A4-B2-C1, i.e., address 101 of the first neuron (highlighted in Figure 11). In the second neuron, WiSARD will write a number 1 in the position indicated by pixels A1-C4-A5 = 101. This process is executed for each pair tuple/neuron of the discriminator. At the end, we have an example of a letter E already learned by WiSARD (Figure 11).

[21] A pixel is the smallest unit of a digital image.

[22] A tuple is a finite ordered sequence of elements.

[23] Actually, the draw is pseudo-random, because computers are not able to generate truly random numbers (Budach 1991).

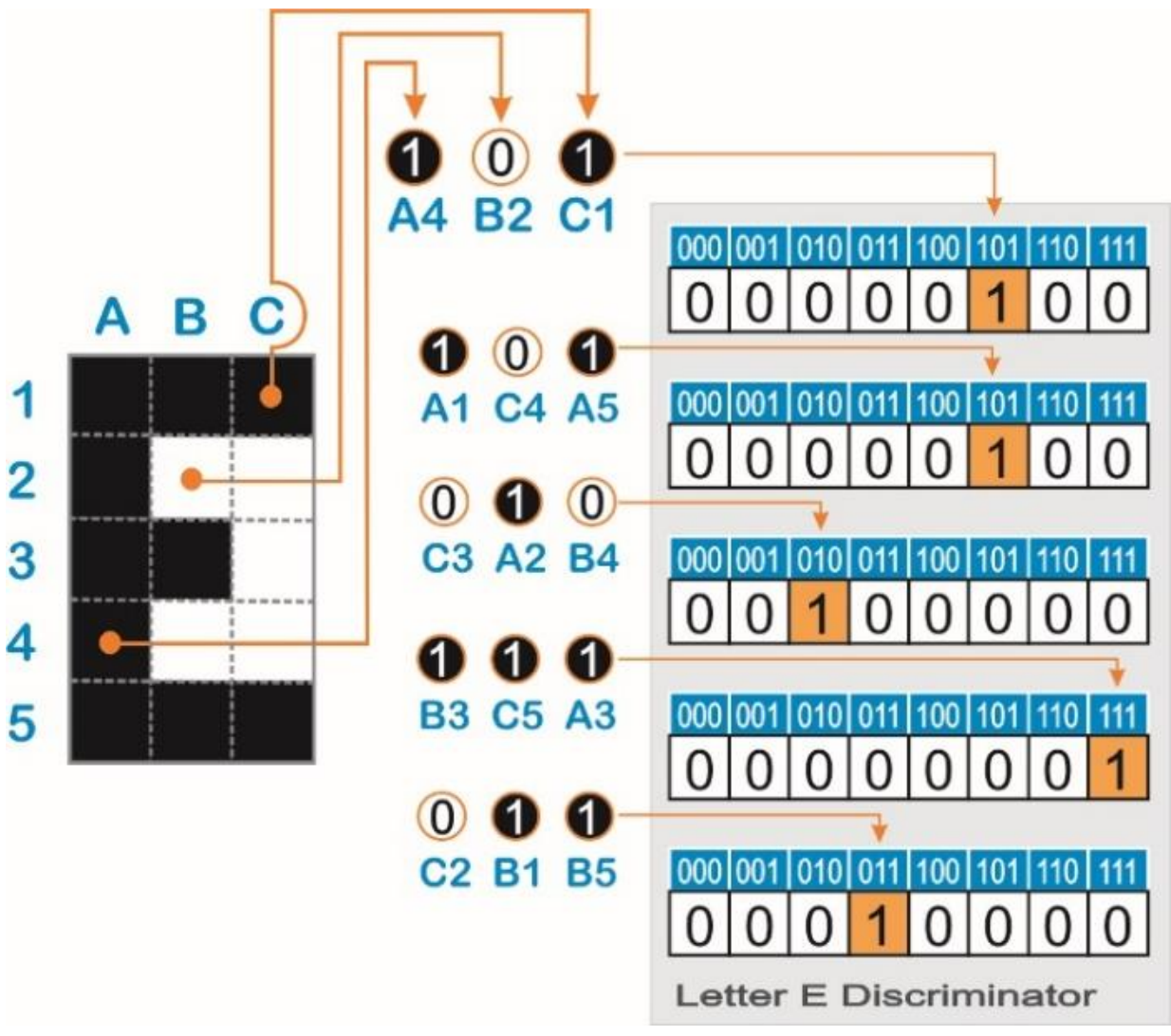

Figure 11: WiSARD learning an example of a letter E

We can then teach a second example of letter E, which must have the same Retina as the first. The Mapping is also the same. After training the second example of the letter E, the Discriminator of the letter E is as shown in Figure 12.

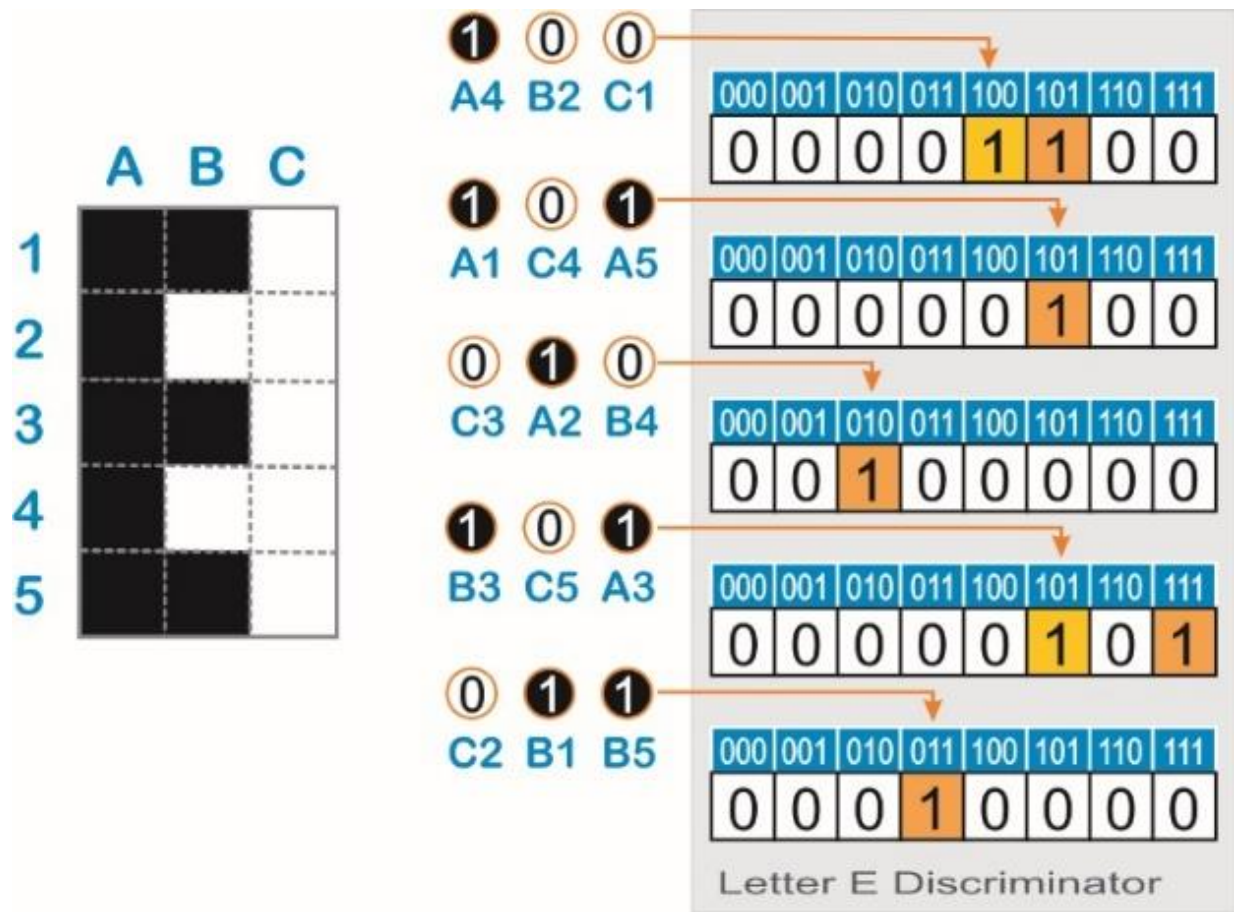

Figure 12: WiSARD learning a second example of a letter E

Now we can check if our WiSARD can recognize a letter E that it has not learned yet, i.e., generalize from what it has already been taught. In the *Classification Phase*, WiSARD checks what value is written at the accessed position of a neuron. That is, WiSARD performs a reading at the accessed position instead of writing. If the value read is the number 1, WiSARD registers that that neuron had an output = 1. Otherwise, if there is a 0 at the accessed position, WiSARD registers that the neuron had an output = 0 (Figure 13). Note that if we present one of the two examples of letter E already trained for WiSARD to recognize, all neurons will display output 1. That happens because the neuron positions to be accessed will be the same as in the Training Phase. So, let us test the classification of a letter E example not taught to WiSARD (Figure 13).

To perform the classification, WiSARD uses an *Adder*. The Adder provides a measure of similarity. It indicates a measure of pertinence of the input to the class of data a Discriminator was trained to recognize. Summing all outputs = 1 resulting from the classification process presented in Figure 13, we get a result = 4. Four of the five neurons had output = 1, which means that the pattern presented for WiSARD is quite similar to the set of examples of letters E taught to it.

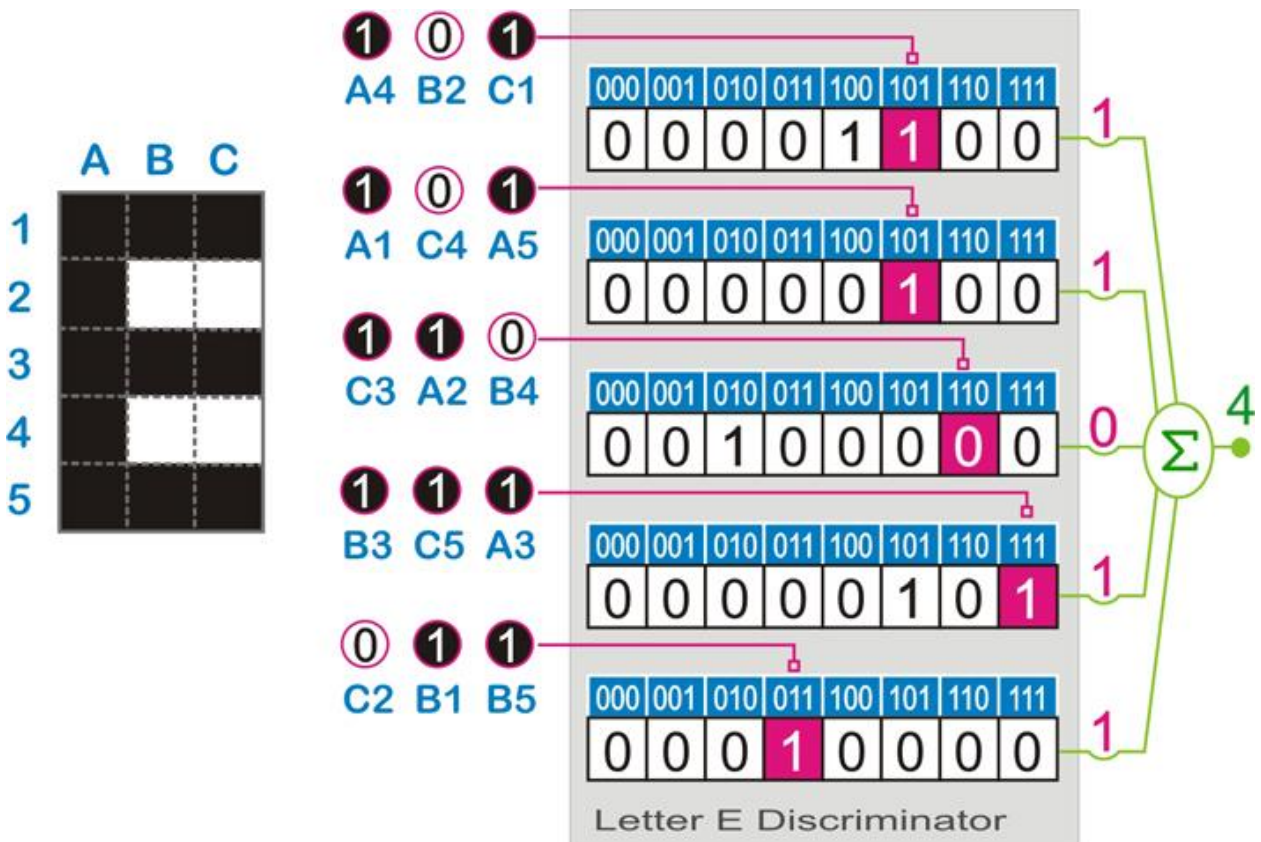

Figure 13: WiSARD classification with only one discriminator

Let us now train a new Discriminator with two examples of letter T (Figure 14 and Figure 15), performing the same process adopted for teaching the letters E. To learn a new class (in this case, a new letter), WiSARD creates a new Discriminator with 0s in all the positions of neurons. For all new examples of letters taught to it, WiSARD will use the same mapping employed to learn the first letter class (E). This is usually the case, but there could be a different mapping for each Discriminator in a multi-discriminator WiSARD.

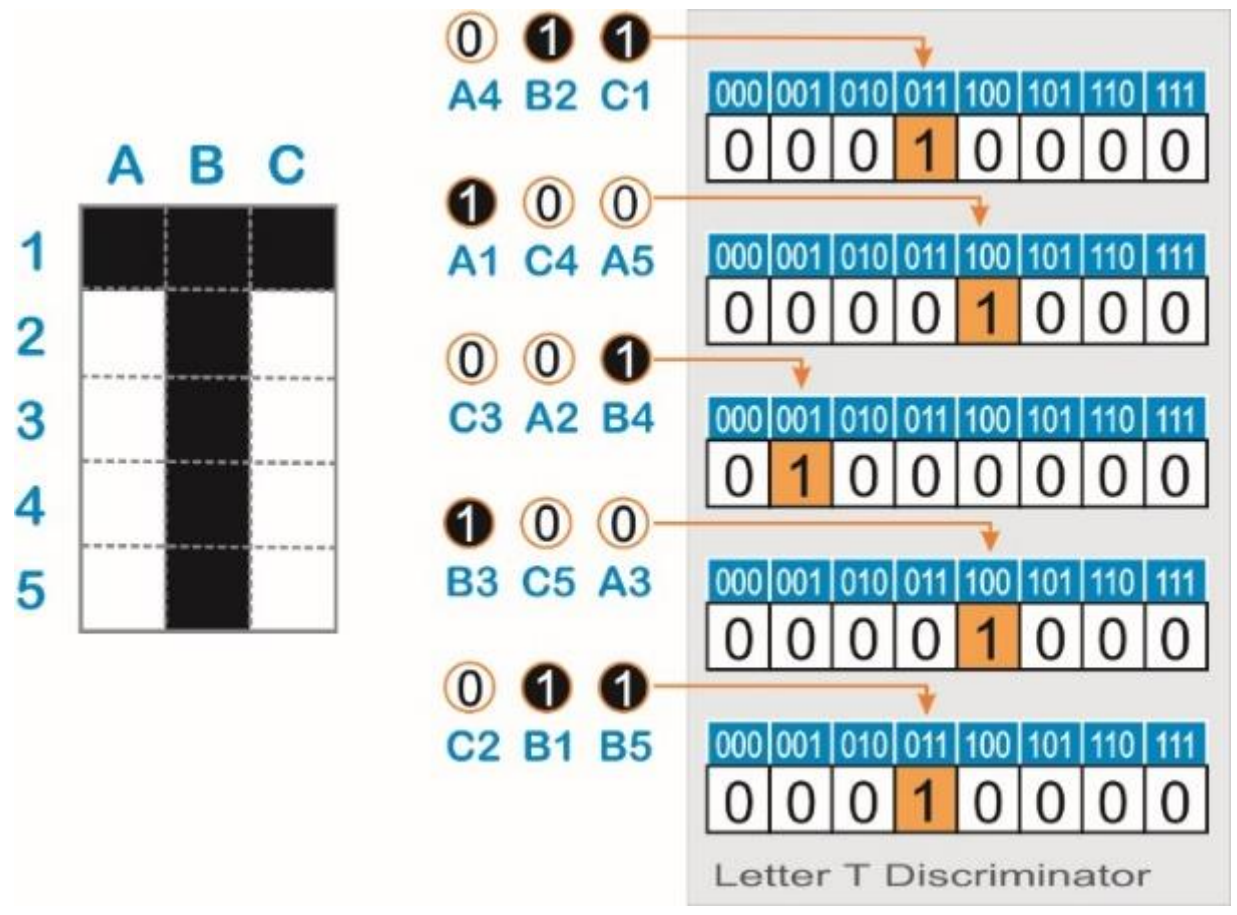

Figure 14: WiSARD learning the first example of a letter T

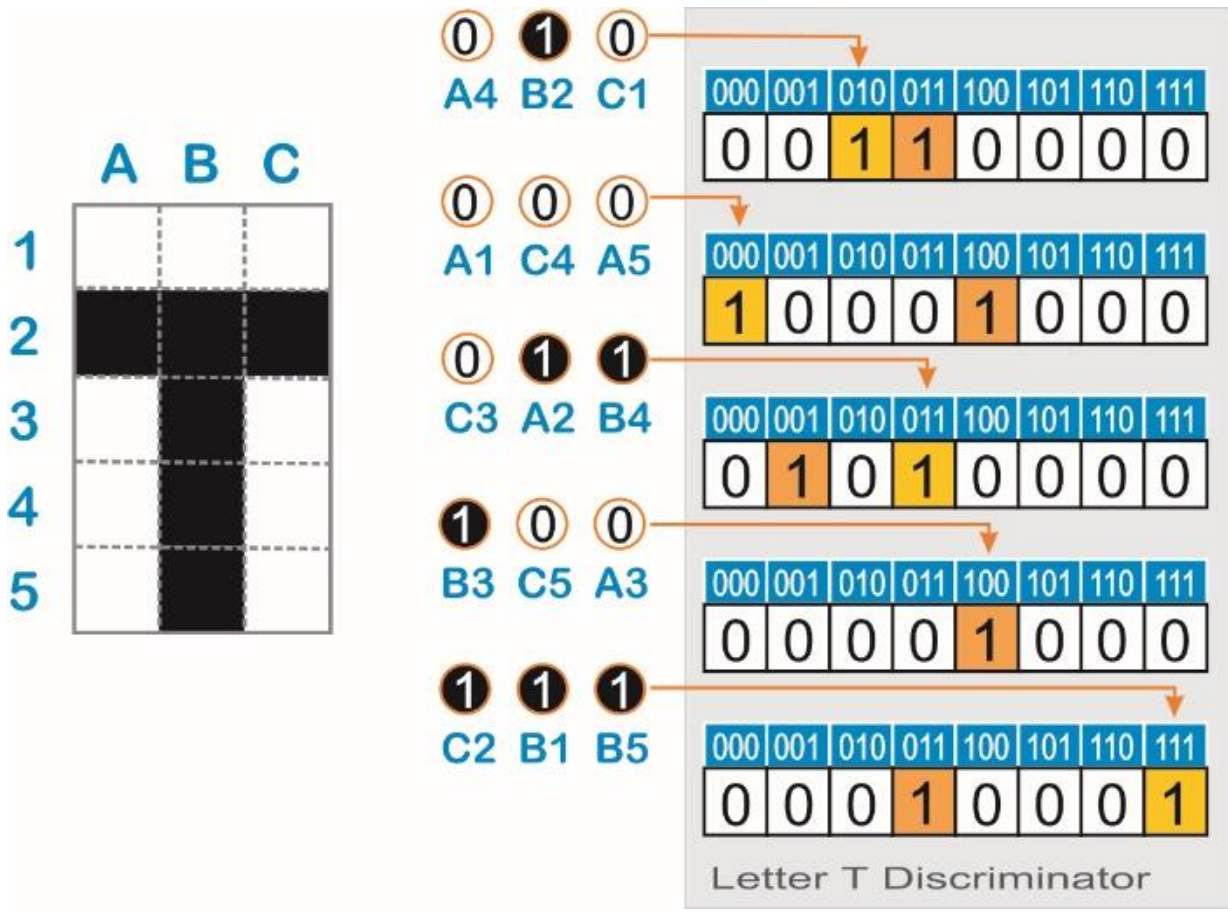

Figure 15: WiSARD learning a second example of a letter T

Now WiSARD can be tested to distinguish letters E from T. In the *Classification Phase,* WiSARD maps the pattern presented in its input to all existing Discriminators. Figure 16 shows that, at the end of this process, the Adder of letter E Discriminator presents a higher value than the Adder of letter T Discriminator. This result indicates that the pattern presented to WiSARD is more like an E than a T. In Figure 17, we have the classification process now performed for an example of a letter T not taught to WiSARD yet. This time, the Adder of letter T Discriminator gives a higher result, and WiSARD reports that the letter pattern presented to it is probably a T.

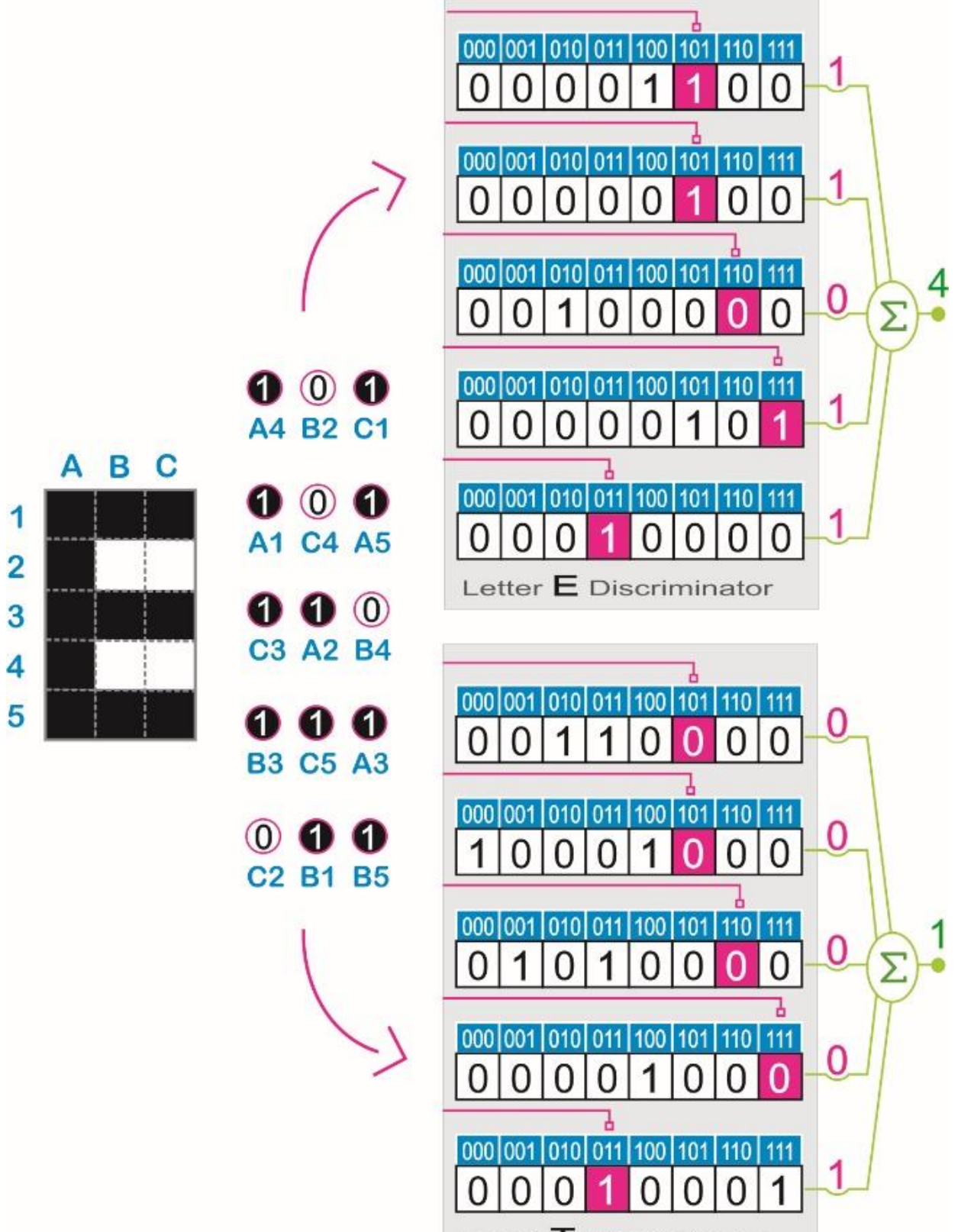

Figure 16: A new observation (more E like) presented to both E and T discriminators

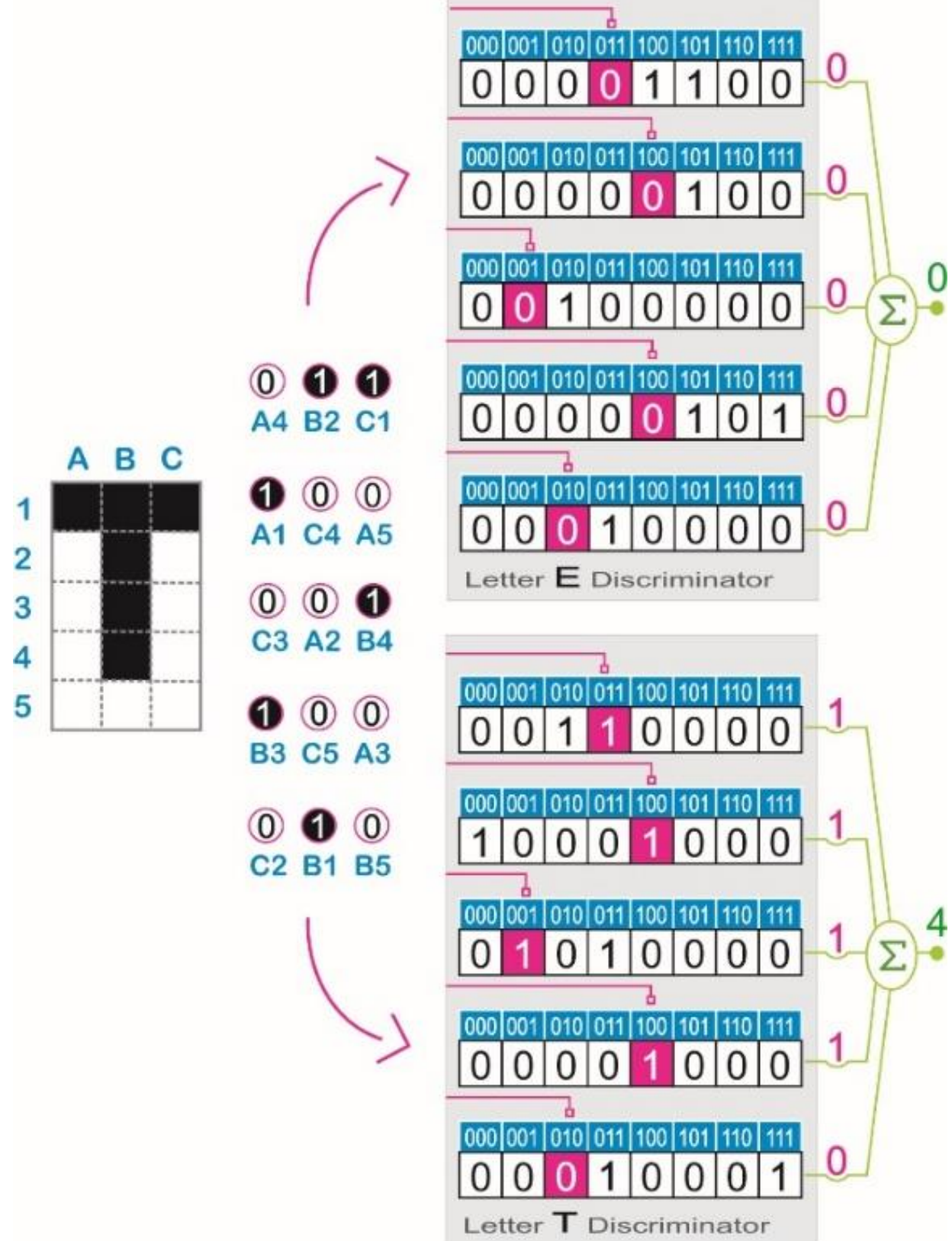

Figure 17: A new observation (more T like) presented to both E and T discriminators

We can observe that all the mechanisms presented here are simple and can be replicated through ludic activities. This makes WiSARD a suitable model for presenting an example of a machine learning process to the general public. The understanding of this process works as a base for building a series of debates that contribute to the demystification of AI, and to developing a sharper perception of its impact on today's society. Furthermore, the *lightweight* of WiSARD allows its easy integration into a block-based programming environment embedded in a low-cost computer. In this way, the benefits of this learning can be carried to more people once the system can be used by low-income or physically isolated communities with hard access to internet services.

### 4.3. WiSARD with bleaching

In original WiSARD, all positions of the neurons are initialized with 0, and a value 1 is written in the positions accessed during the training phase. When trained with a large number of examples for each class, discriminators can become saturated, i.e., with many RAM positions containing the value 1. Thus, for a given observation, a tie may occur between discriminators. In this case, WiSARD needs to randomly choose one of the classes from which the tie occurred to present it as an answer. The *bleaching* technique was developed to reduce this problem by including a tiebreak mechanism (Grieco et al. 2010; Carvalho et al. 2013).

For that, WiSARD training algorithm is modified in the following aspect: instead of storing only the value 1 in the positions accessed in the neurons, this value is increased by 1 each time a given position is accessed. The values stored in the neurons of a discriminator will vary between 0 and the number of patterns used for training (Grieco et al. 2010). In the original WiSARD, during the classification phase, a neuron's output is equal to 1 when the accessed position has a value 1 stored. Using the bleaching technique, the output of a neuron is 1 only if the value accessed is greater than or equal to a bleaching value $b$ (Franca et al. 2014).

The value of $b$ starts at 1. In this situation, the outputs of the neurons are equal to the outputs of an original WiSARD. If a draw is observed between the responses of the discriminators, $b$ is incremented (usually by 1) until there are no more ties. The class whose discriminator has the highest output is the chosen one (Franca et al. 2014, Carvalho et al 2013). Figure 19 shows the output of a discriminator trained with the examples of Figure 18 for a new observation having $b$=2.

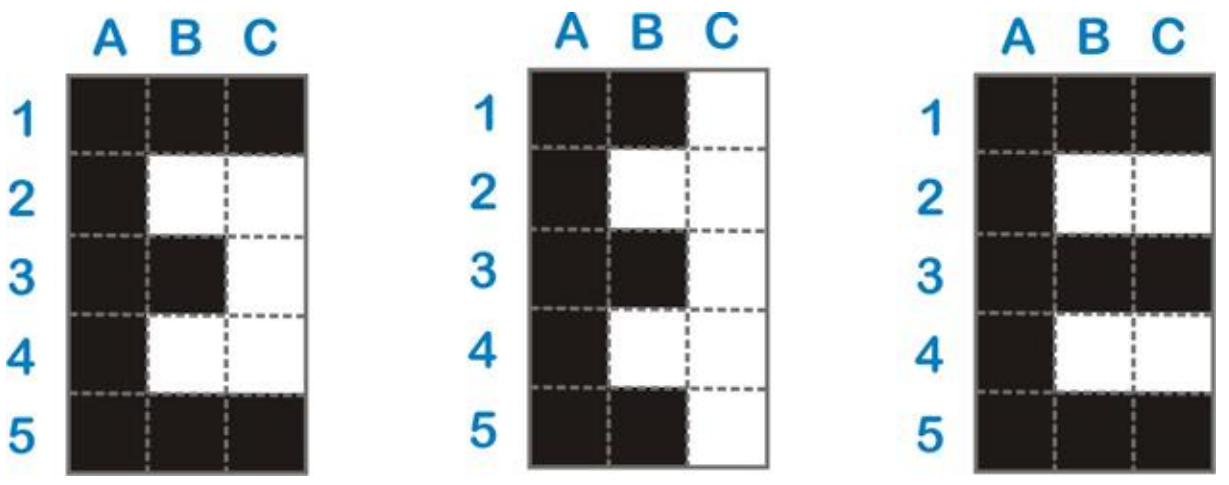

Figure 18: Examples used to train the letter E discriminator

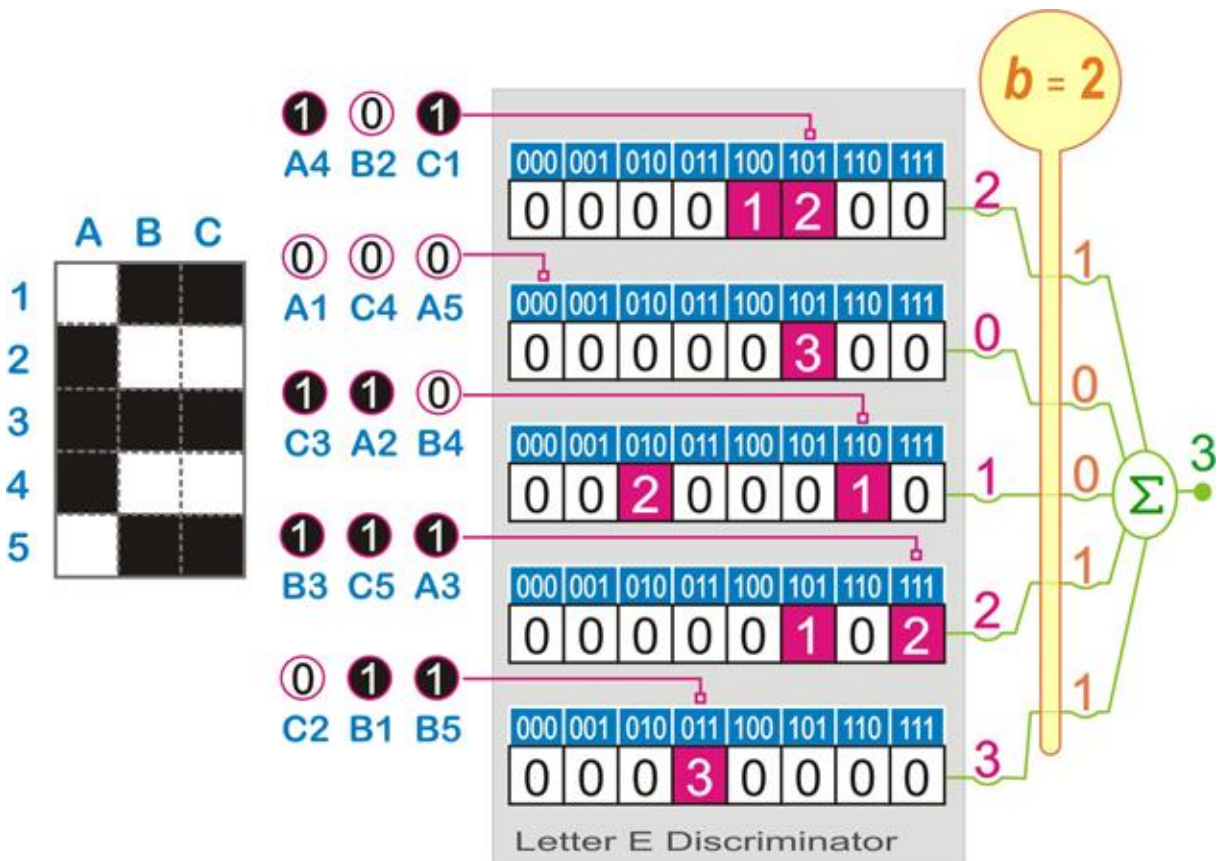

Figure 19: An example of the letter E discriminator output for an observation with bleaching $b$ = 2

## 4.4. DRASiW and mental images

DRASiW is an extension of WiSARD that can produce a representative example of a class learned from trained patterns. This is based on a WiSARD with bleaching, presented in Subsection 4.3. The values stored in the neurons are used as counters that can be reversed in an internal retina where a *mental image* is produced. As seen before, in the classification phase, WiSARD returns the name of the class to which a given observation belongs. Using DRASiW, one can present a class name to WiSARD and it returns a prototype of the patterns used to train that class (Grieco et al. 2010).

To extract a mental image from a discriminator, an internal retina with the same dimensions as the retina used to train WiSARD is created. Then, the content of each position of the discriminator's neurons is added to the coordinates of the internal retina corresponding to bits 1 of the neuron's position address. The mapping used is the same. Once this process is finished, a gray level can be associated with each coordinate of the internal retina. The shades of gray will be proportional to the values present in the coordinates. The highest value will correspond to the black color, and the value 0 will correspond to the white color (Grieco et al. 2010). Figure 20 shows this procedure being performed only on neuron positions with non-zero content for the discriminator of letter E (see Figure 19).

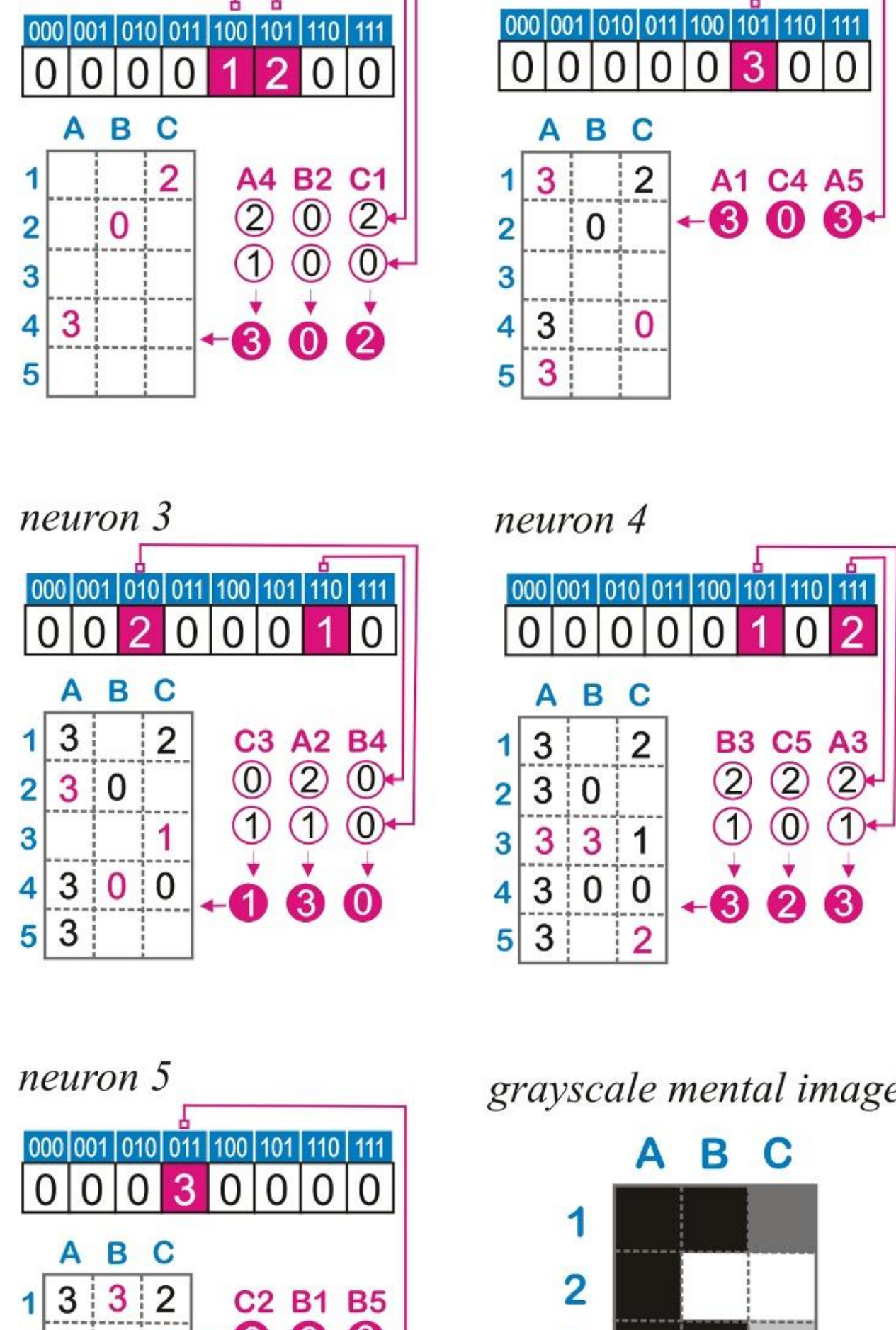

Figure 20: Mental image extraction for class E

## 5. AIcon2abs: AI from concrete to abstract methodology

AI literacy enables people to understand concepts and methods related to artificial intelligence instead of just learning how to use systems and devices that include techniques belonging to this field of knowledge (Kandlhofer et al. 2016). In this section we will present how the combination of block-based programming, WiSARD, and educational robotics, facilitates the understanding of some basic concepts of machine learning and AI to a broader audience. We also describe how Constructivist and Constructionist theories, the agent-based approach, the machine learning process as described by Mitchell (1997), and the ideas proposed by Turing (1950) through the Imitation Game, support AIcon2abs methodology.

### 5.1. BlockWiSARD design elements

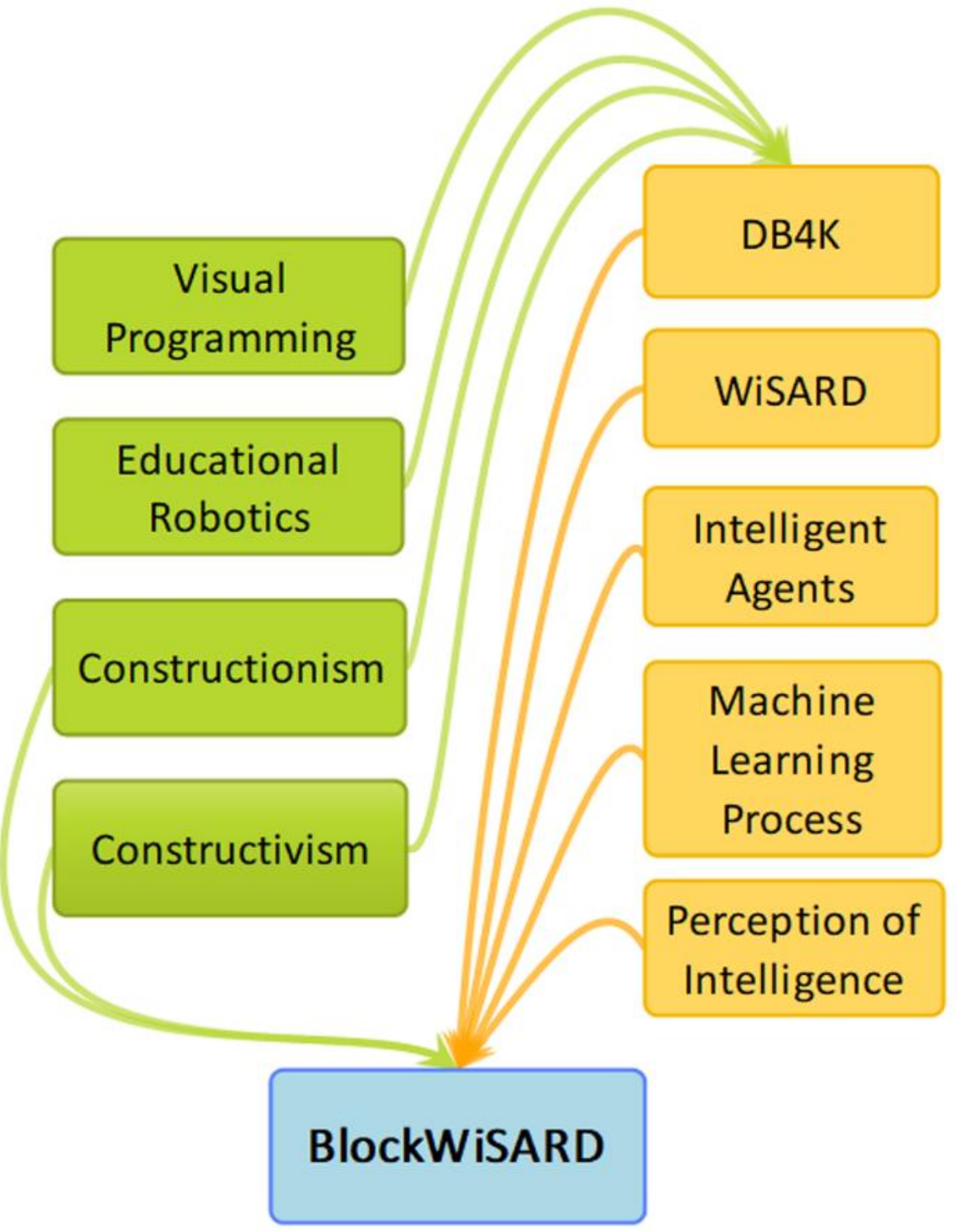

Figure 21: Concepts and technologies involved in the development of BlockWiSARD

One of the obstacles to help the general public to develop a basic understanding of artificial intelligence is its complexity and the background knowledge necessary to be previously acquired (Sakulkueakulsuk et al. 2018). BlockWiSARD was developed to overcome this initial barrier. It consists of a visual programming environment based on snap blocks that makes use of WiSARD WANN to enable general people to develop systems with some learning capability. Figure 21 shows de concepts and technologies involved in the development of BlockWiSARD.

From DB4K, BlockWiSARD inherited both block-based programming and educational robotics concepts. By using WiSARD, BlockWiSARD incorporates the possibility of building systems that learn from examples. The difference of a computer system capable of learning from a conventional computer system is observed from the agent approach and the machine learning process concept. Finally, the verification of the existence or not of *intelligence* in the developed systems is built from the learner's perception of the behavior presented by these systems. The elements added to the development of BlockWiSARD have constructionist and constructivist characteristics that help by reducing the abstraction capacity required for people to learn.

### 5.2. BlockWiSARD environment

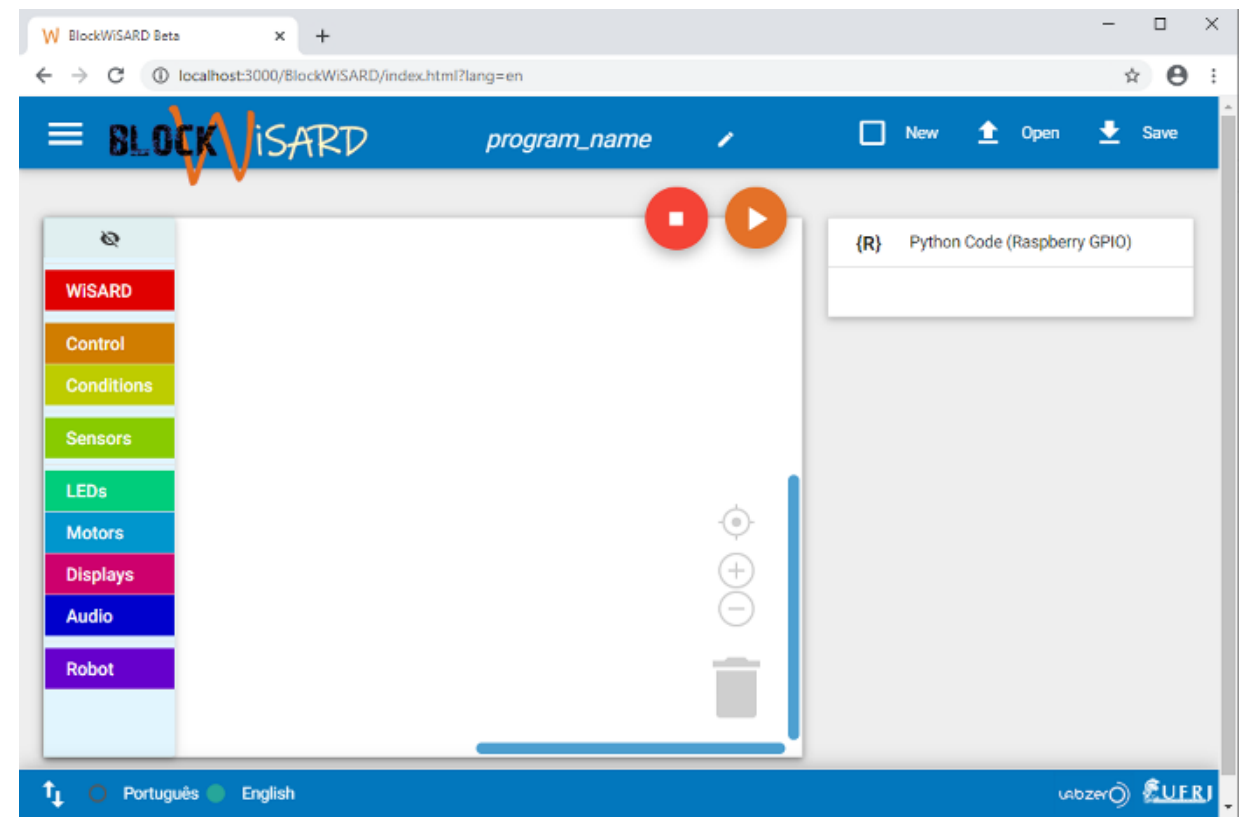

Figure 22: BlockWiSARD interface overview

The main idea behind BlockWiSARD is simplicity. The programming environment provides a little set of low complexity components with a semantics based on observable aspects of the concepts to be addressed. The environment's interface (Figure 22) has two main elements: the *Toolbox* (Figure 23), where the user picks the blocks, and the *Workspace* (Figure 24), where the user places the blocks to create the programs.

The Workspace (Figure 24) has a set of controls for zooming the block program in and out, a trash can, and scrollbars. These are original elements provided by Blockly[24]. Additionally, we have an orange button on the top of the workspace responsible for running the Python[25] code generated by the block program, and a red button used to stop the running program. The main window also has interface elements to save a block program, to load a saved block program, to save the python code generated by the block program and to open, in python IDLE[26], the python program corresponding to the block program present on the workspace. At the right side of the screen, the environment has an area to display the python code corresponding to the block program. The Toolbox (Figure 23) is organized in 5 groups of blocks: WiSARD, Controls + Conditions, Sensors, Actuators (LEDs, Motors, Displays and Audio), and Robot.

[24] Blockly is the library used to develop BlockWiSARD. https://developers.google.com/blockly

[25] Python is a programming language https://www.python.org/. BlockWiSARD converts the block program created by the user into a program in that programming language.

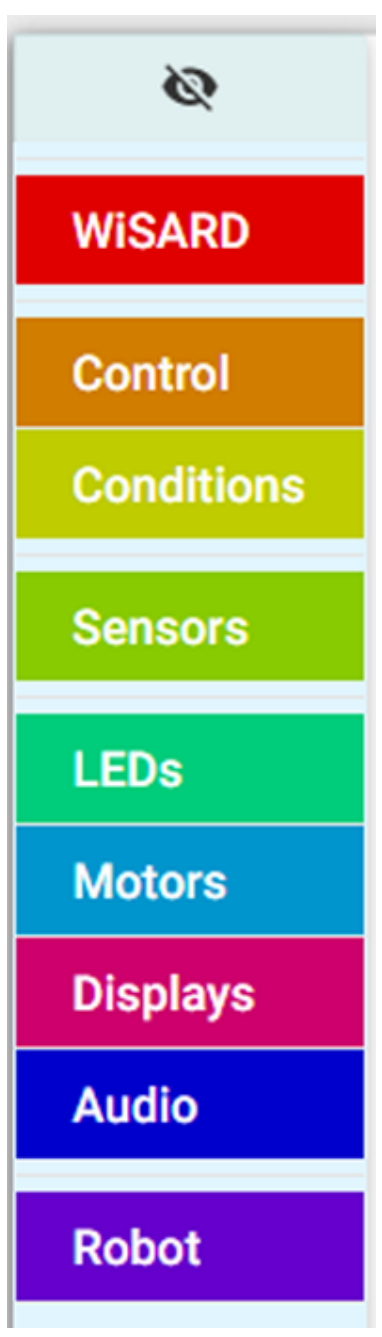

Figure 23: Toolbox

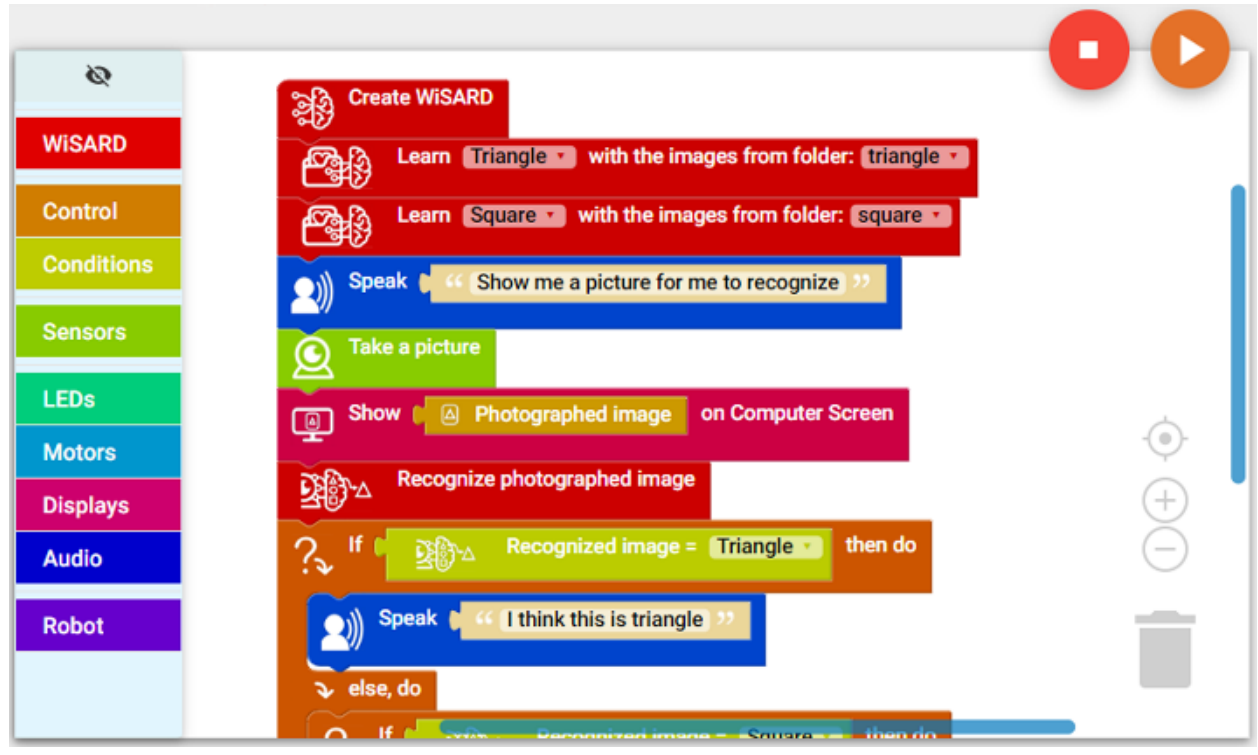

Figure 24: Workspace with an example program

The Control blocks group (Figure 25) is responsible for the program flow control. *Conditional Repetition Structure* and the *Decision Structures* use blocks presented in the Conditions Group (Figure 26) as conditioning factors. In turn, *conditioning* blocks use values obtained by sensors as reference values for verifying the satisfaction of imposed conditions. These values can be acquired directly through the sensor group blocks or from the *Recognize Photographed Image* block present in WiSARD group.

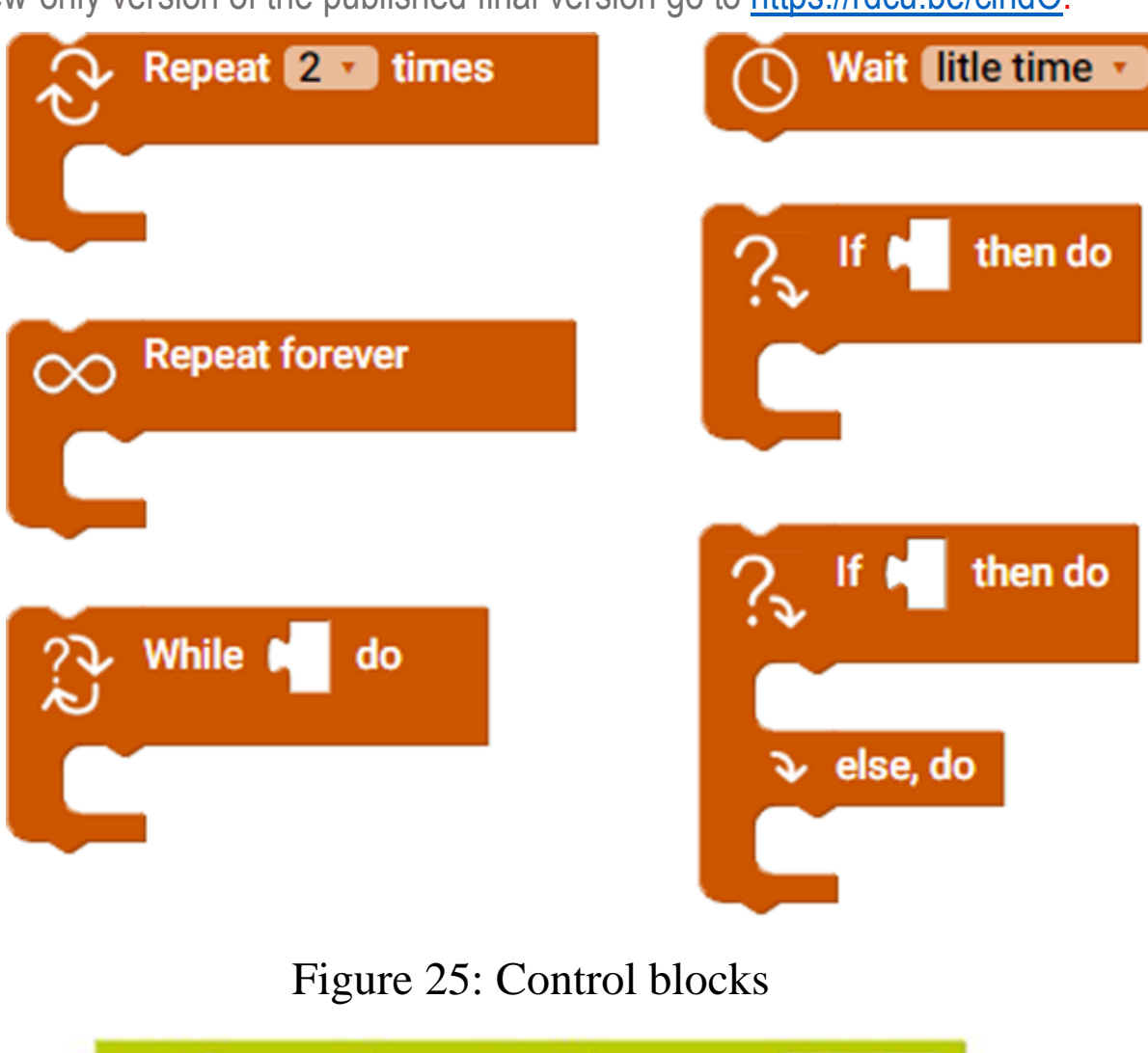

Figure 25: Control blocks

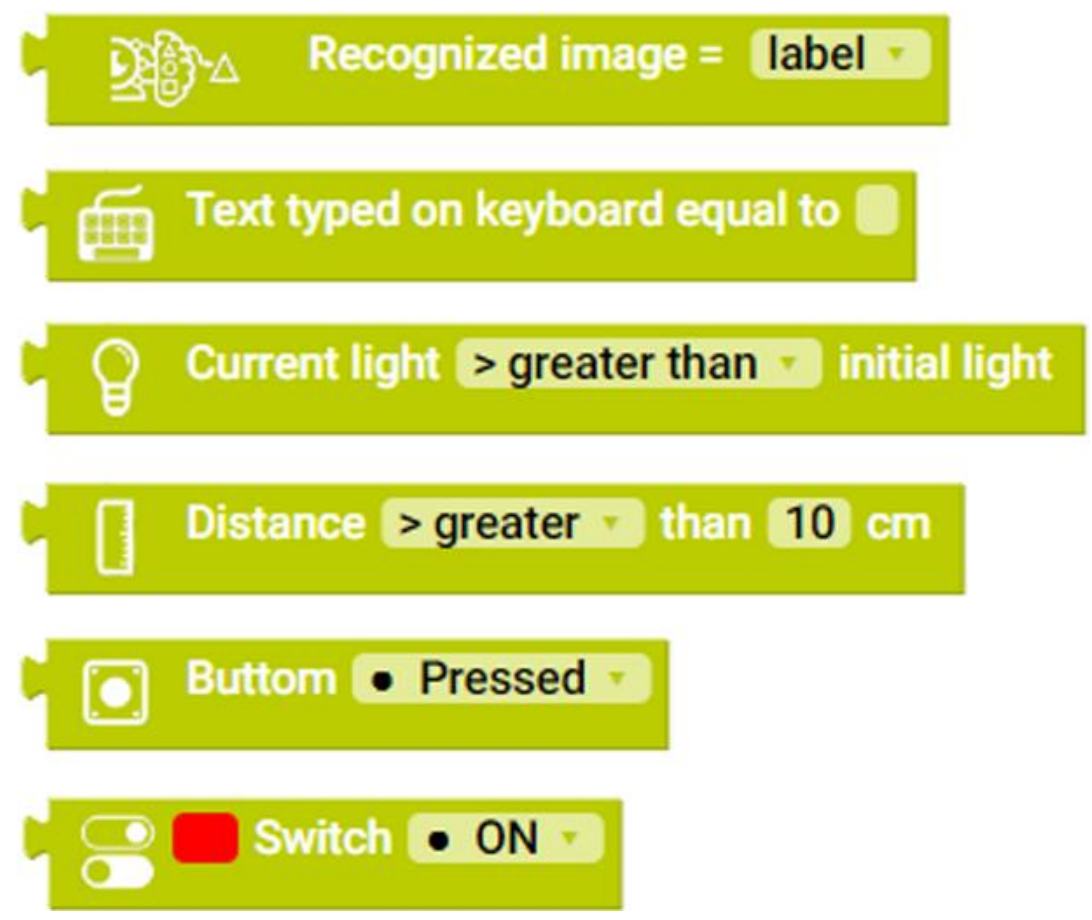

Figure 26: Conditions blocks

Sensors blocks group (Figure 27) has a block to take pictures with the webcam, a block to get data from the keyboard, a block to acquire distance values, two blocks for data acquisition regarding luminosity, and blocks for reading the state (on/off) of a pushbutton and switches. The golden blocks are used to present data read by the sensors. They are used in combination with the blocks *Show < picture > on Computer Screen*, *Write < text > on the Computer Screen*, *Write < text > on 16x2 LCD line < number >*, and *Speak < text >* (Figure 30 and Figure 31).

Actuators blocks are grouped in four sets: (LEDs (Figure 28), Motors (Figure 29), Displays (Figure 30) and Audio (Figure 31). By using these blocks, the programmer can control the devices most commonly used in educational robotics activities, write on the computer screen, play some sound effects, and use the text-to-speech functionality. The last group is the Robot blocks group (Figure 32). These blocks are used to control the movements of a pair of wheels attached to a chassis.

WiSARD group is responsible for providing the systems developed with BlockWiSARD with the ability to learn. Next section details this group.

[26] IDLE is Python's Integrated Development and Learning Environment. https://docs.python.org/3/library/idle.html

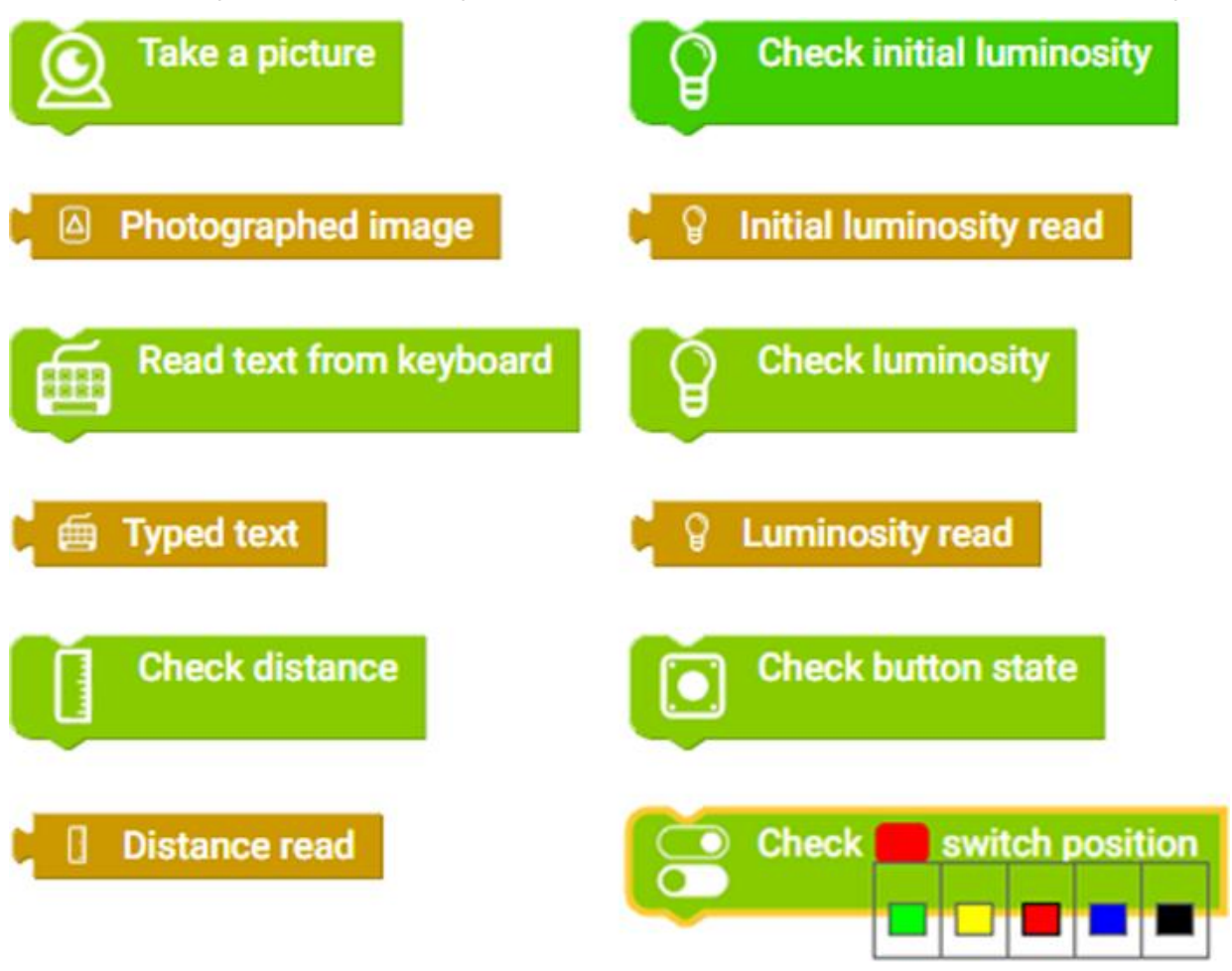

Figure 27: Sensors blocks

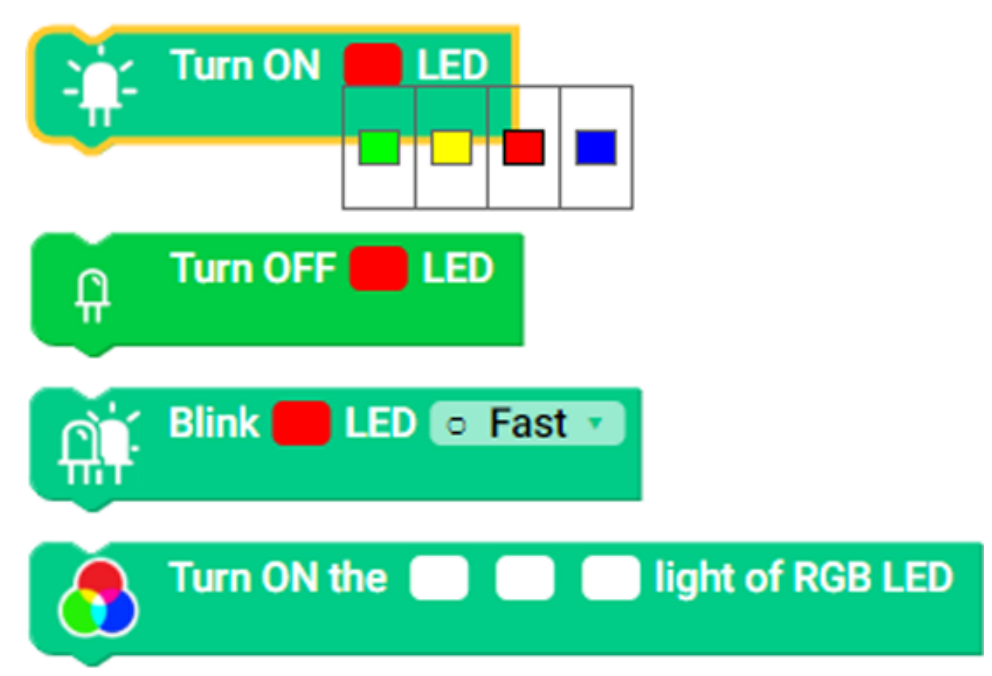

Figure 28: LEDs blocks

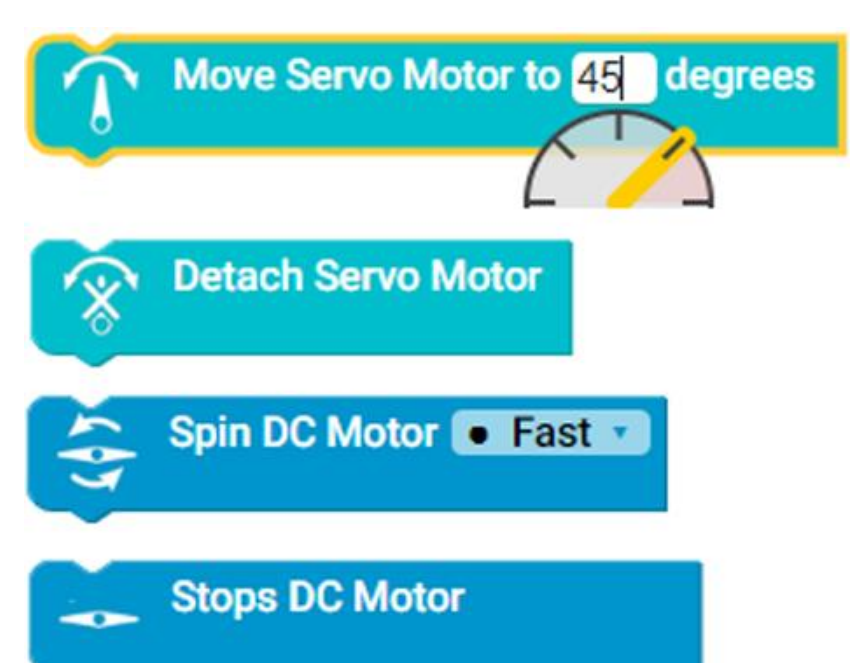

Figure 29: Motors blocks

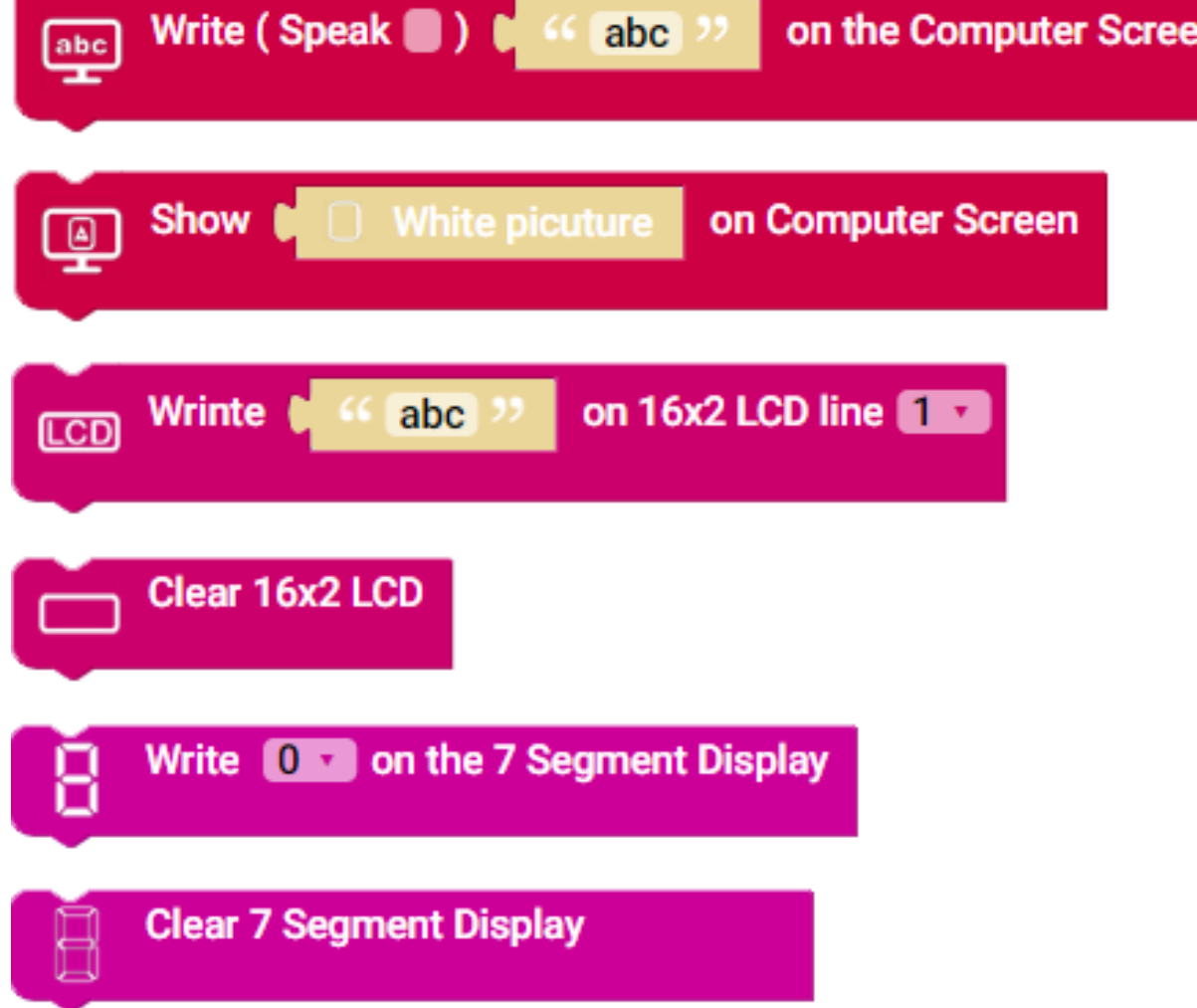

Figure 30: Displays blocks

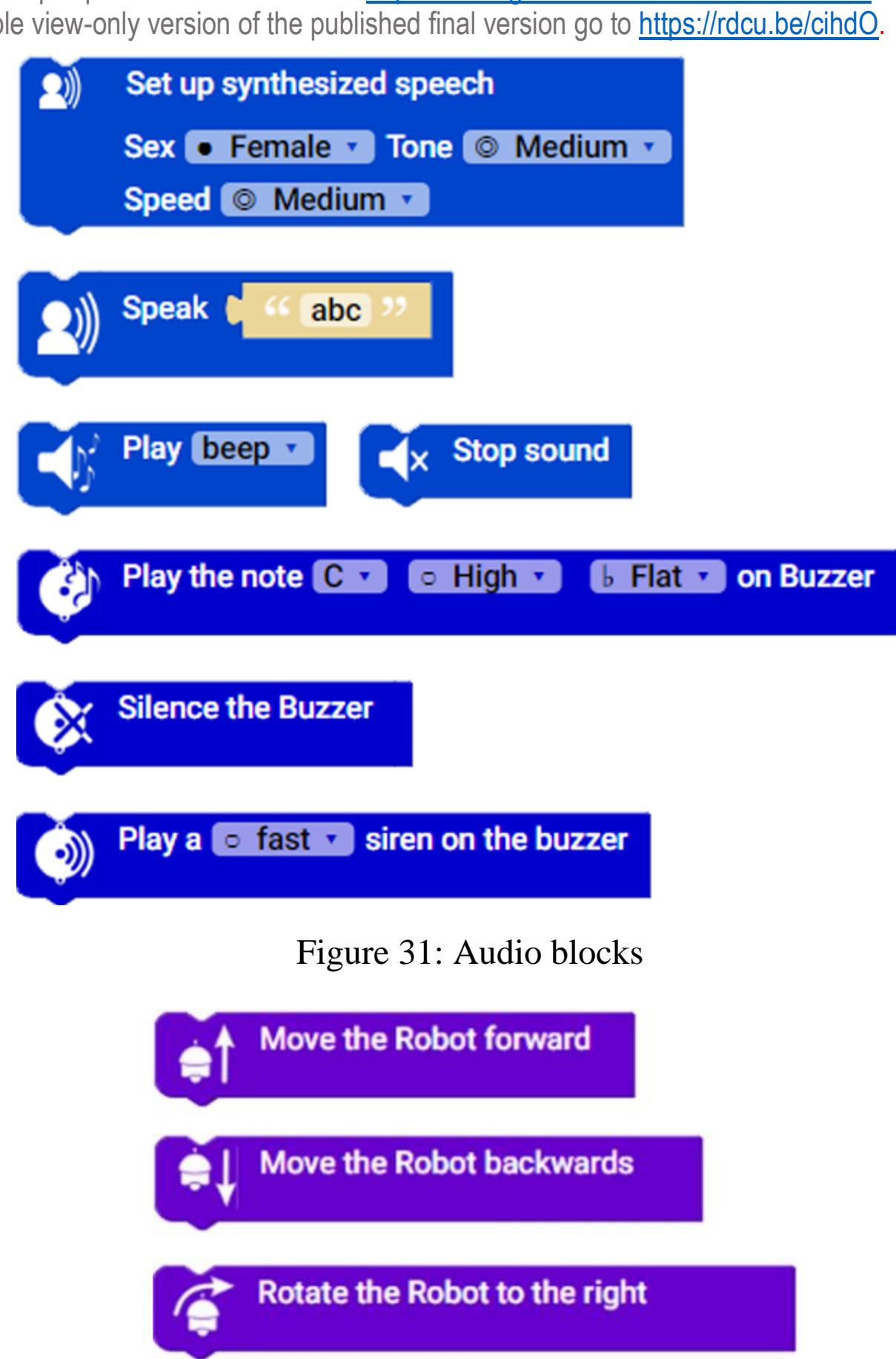

Figure 31: Audio blocks

Figure 32: Robot blocks

## 5.3. WiSARD blocks

In a conventional computer program, the programmer *says* to the computer the meaning of all data it is going to use to make decisions. For this reason, we know exactly how it will *act* when fed with certain data. Using AI is different, the computer builds the meaning of a set of data presented to it (with human help in a supervised learning approach) and then determines, by itself, the meaning of new data based on what it has learned before. As a result, we do not know (*a priori*) what the computer will do (within a set of possible predetermined actions) when fed with certain data. This decision will depend on the meaning that the machine is going to attribute to new data presented to it. A way to evidentiate this fundamental difference is to include the data acquisition, training, and classification primitives as elements of an otherwise conventional computer program.

Another tool adopted by BlockWiSARD to highlight that distinction is the adoption of the agent's approach. The box with the question mark, present in Figure 33, represents the agent's decision mechanism. The agent uses it to say to its actuators what to do in response to the data acquired through its sensors. In non-intelligent systems, as in the case of de programs developed with DB4K, this mechanism corresponds to a simple comparison of the input data with values pre-defined in the computer program code. With BlockWiSARD, it is possible to turn the black box into an intelligent one, capable of learning and generalizing from what was learned. The facility of developing both non-intelligent and basic intelligent agents helps to clarify the differences between a conventional computer program from a system able to learn.

Tom Mitchell's (1997) definition of the machine learning process was also adopted in BlockWiSARD: we can say that an agent is learning when its performance *P* in executing a task *T* improves with experience *E*. The observation of these variables helps to evidentiate, in a simple manner, the perception of the learning process of the agent. In BlockWiSARD we have:

- Task *T*: To recognize pictures drawn in white papers.
- Performance measure *P*: The number of pictures correctly recognized. This can be observed through the actions performed by the agent in response to the classification process.
- Training experience *E*: A database of pictures with respective labels presented to the agent in a *batch* process and/or in an *online* learning fashion. In batch learning, training and classification take place separately. The system needs to be trained with the entire dataset before executing any classification task. To incorporate new data, it is necessary to retrain the system with the whole dataset that includes the new data. In online learning, training is interleaved with classification.

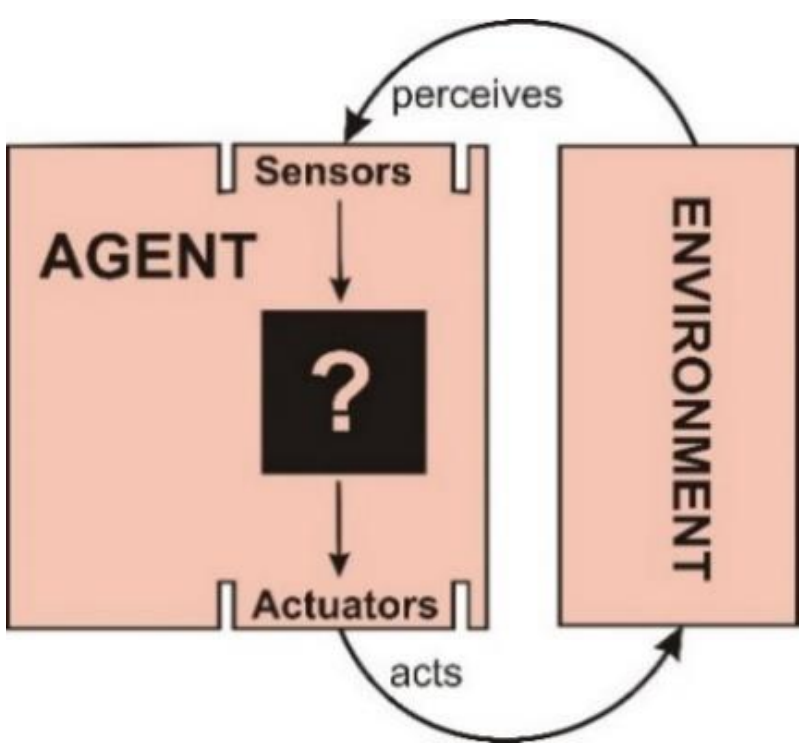

Figure 33: An agent interacts with the environment through sensors and actuators (Russel and Norvig 2010)

WiSARD WANN is ideal for implementing activities around this concept because it can learn entirely online quite fast and with few examples. This characteristic helps to make it easily observable the gain in performance *P* on task *T* as training experience *E* increases.

To include the ability to learn in the programs developed with BlockWiSARD, we designed four blocks directly related to the process of training and classification (Figure 34).

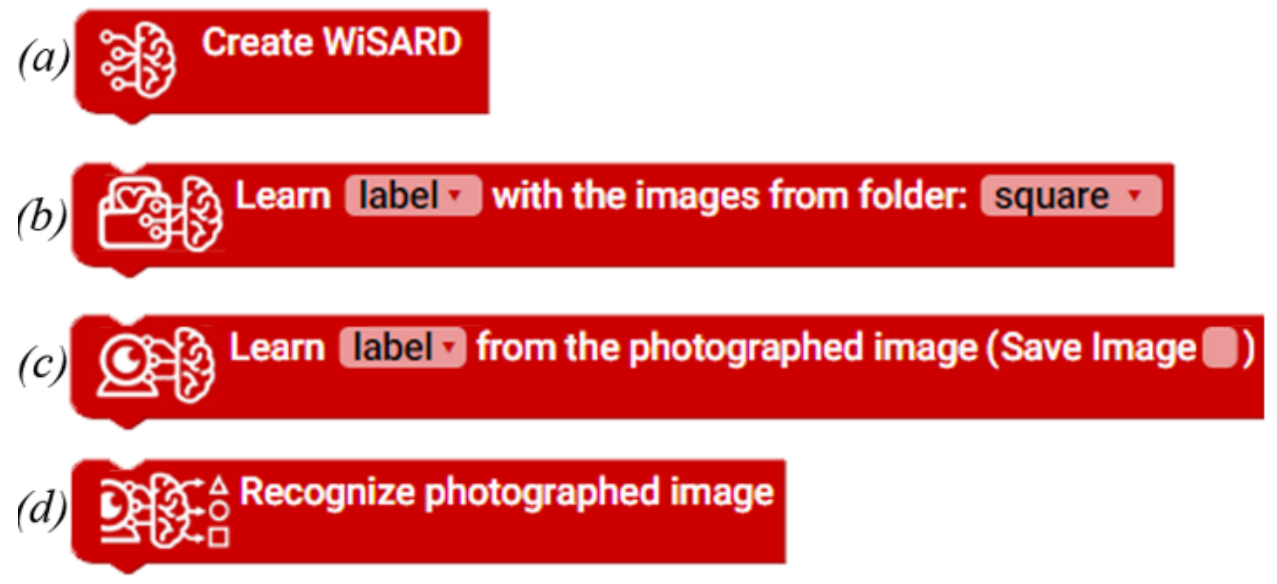

Figure 34: BlockWiSARD machine learning blocks. a) Create WiSARD instance; b) Learn from a dataset of images; c) Learn a photographed image; d) Recognize photographed image

By combining this set of blocks with the blocks presented in Subsection 5.2, it is easy to develop systems that:

- create an instance of a learning mechanism
- request data to be learned
- request corresponding labels for that data
- classify new data given as input

The block *Create WiSARD* (Figure 34*a*), *creates* an instance of a WiSARD WANN (BlockWiSARD adopts a WiSARD model with bleaching mechanism). WiSARD is presented to the user as a *kind of brain*. By using this block in the program, the student is providing the machine with the ability to learn. In order to maintain the simplicity of the methodology, that brain can learn only one kind of thing: Identifying black drawings on white background. Also, for the sake of simplicity, each program can have only one brain. So, there can be only one *Create WiSARD* block on the workspace. The size of the retina and the size of the tuple are predefined in a configuration file, being transparent to the user.

Since the machine is equipped with learning ability, the user can include blocks in the program that enable the device to feed its brain with *things* to be learned. There are two blocks with this functionality. The first indicates a folder in the machine with pictures to be learned (Figure 34*b*). When the program is run, each image accessed for teaching the computer is presented to the user on the computer screen. The second block gives the machine the ability to learn from pictures taken in real-time with a webcam (Figure 34*c*). In both cases, the program needs to inform the machine what the picture is. For this purpose, these blocks have a parameter to indicate a label (the class name) for each image. Once defined, the labels are saved and can be picked from drop-down lists. The next block (Figure 34*d*) is used to give the machine the ability to recognize (classify) an image captured with its webcam.

The environment performs some automatic checks to ensure that, for example, a *learning* block is not positioned before the *Create WiSARD* block, and that a *Recognize photographed image* block is not used before a *Take a picture* block. In this type of situation, improperly used blocks are disabled, painted with a pale color, and display a warning indicating the problem (Figure 35). This is useful to guide the student's first steps in building his/her learning machine.

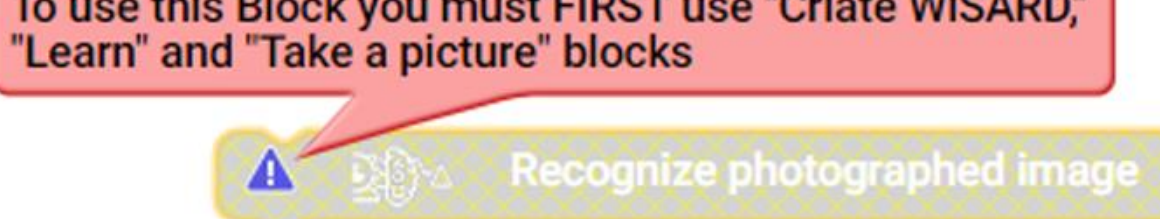

Figure 35: Example of a disabled block

The combination of *Write in the Computer Screen*, *Read Text from the keyboard*, and *Take a Picture,* with *WiSARD* and *Control Blocks*, makes it possible to build a learning machine from scratch, interacting with the user via the computer. By adding the *Robot Blocks*, together with blocks of sensors and actuators, one can create a basic intelligent robotic agent in a quite simple way.

The last one of the WiSARD blocks group is the *Show mental image of* < <u>*label*</u> > (Figure 36). This block generates a mental image of the class indicated by the label parameter. The mental images allow students to see a prototype of a class that has been learnt from trained patterns. Figure 37 and Figure 38 show respectively six dog faces and six flowers used to train BlockWiSARD. Figure 39 shows the corresponding mental images output by WiSARD.

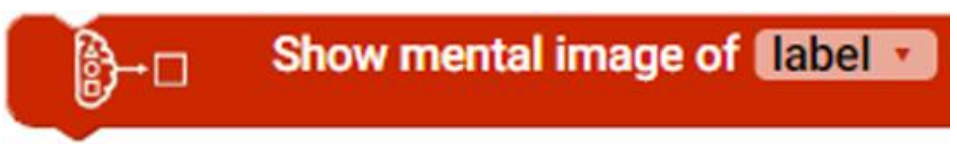

Figure 36: *Show mental image of* < <u>*label*</u> > block

This functionality allows humans to put themselves in the place of the machine and check how they would classify unseen pictures of each pretrained class by comparing them directly with the mental images BlockWiSARD has created. After that, people can show the same untrained pictures to WiSARD and see if it will classify them the same way they did.

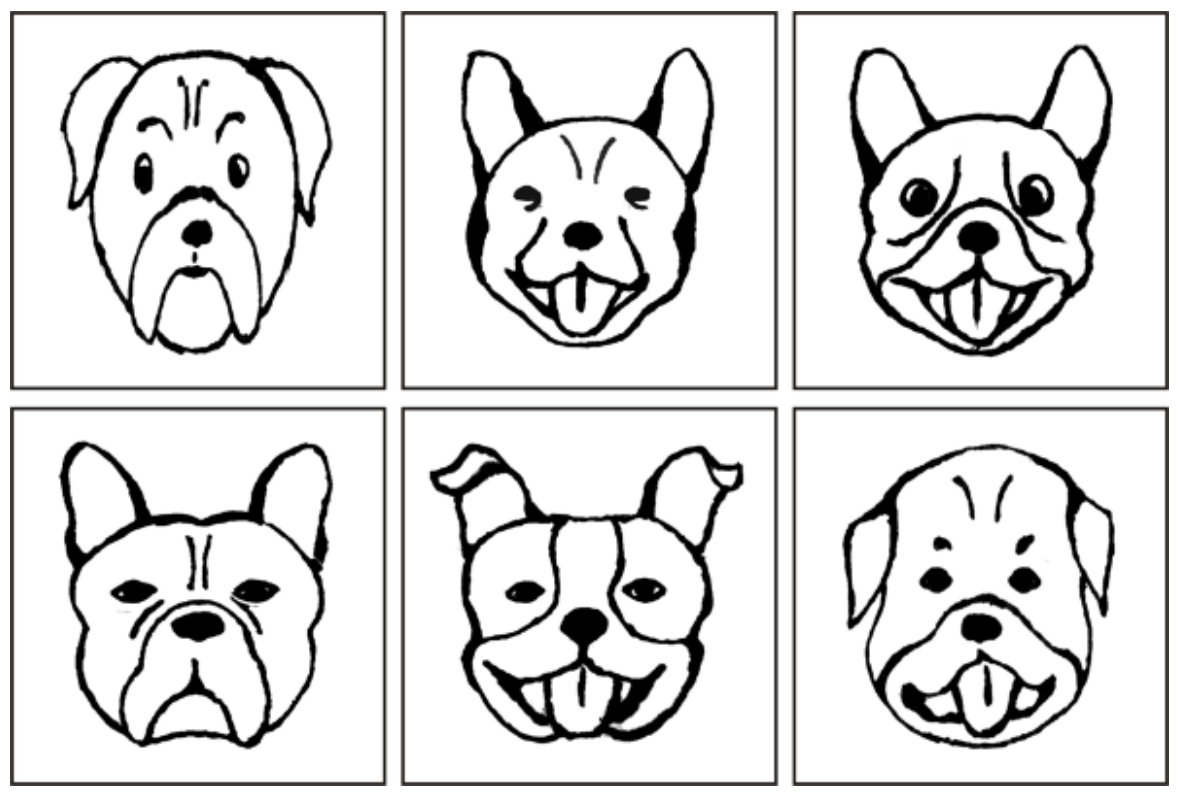

Figure 37: Dog faces used to train WiSARD

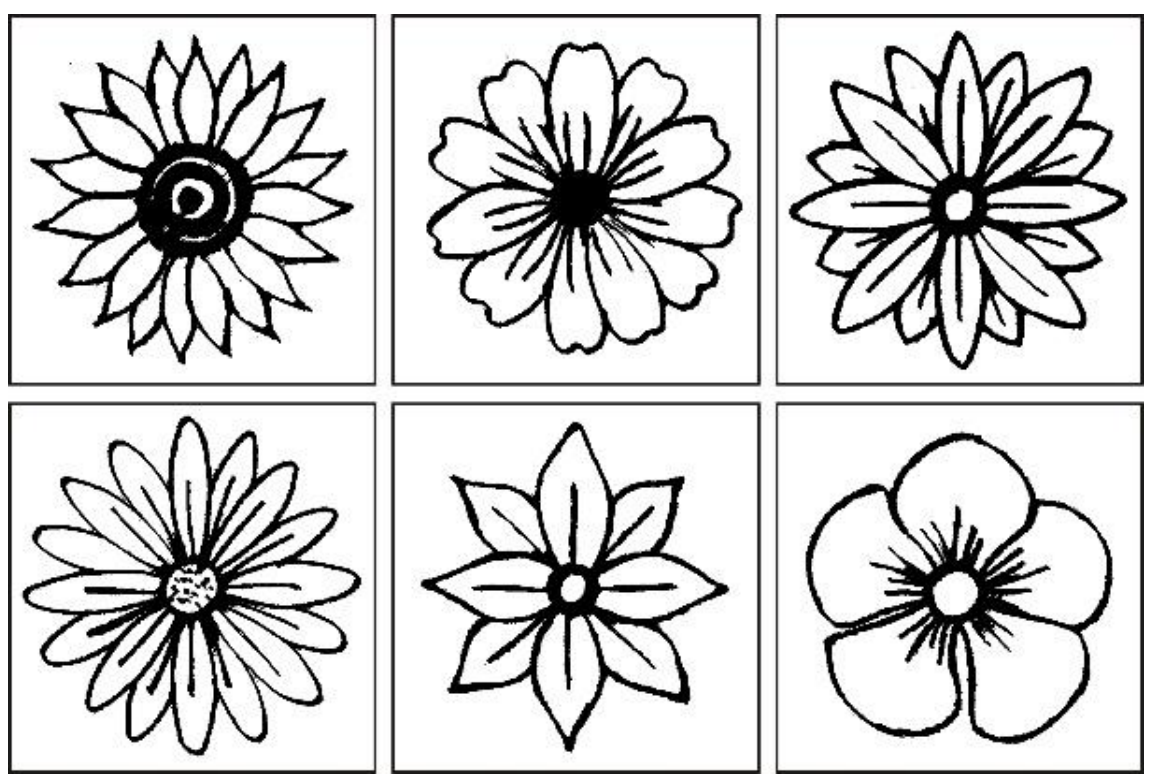

Figure 38: Flowers used to train WiSARD

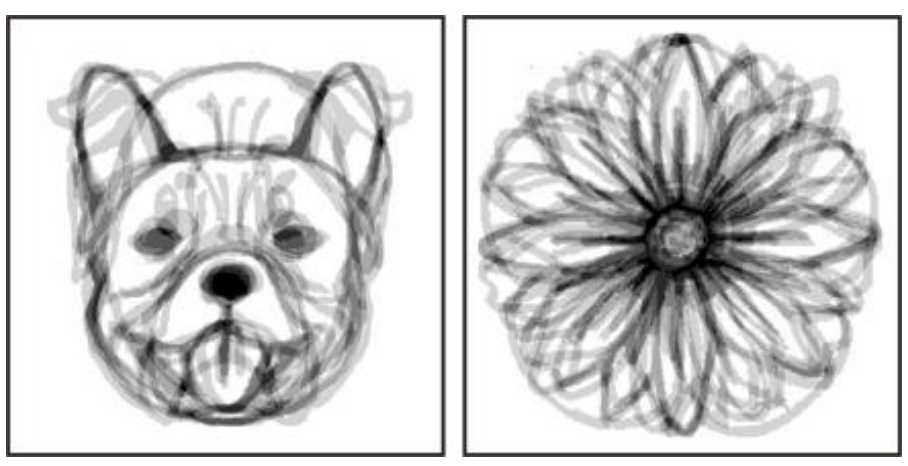

Figure 39: WiSARD mental images of a dog face and a flower

## 5.4. Example of a BlockWiSARD program

Figure 40 presents a program created with BlockWiSARD that *transforms* a computer equipped with a webcam into a basic learning machine. Through this program, the computer can be trained online and classify (generalizing from what has been trained) two different classes of images. The programmer uses the blocks responsible for the machine learning process in the same way as he/she uses the other conventional programming blocks.

When running this program, the computer interacts with the user asking him/her for pictures to learn (train) or recognize (classify). If the user asks the computer to identify any picture before teaching anything to it, the computer tells it does not know what that image is. Teaching the computer only one example of a five petals flower and a five-pointed star (Figure 41) it can generalize from what it has learned and correctly identify the images presented in Figure 42. When presented to the image in the Figure 43 computer confuses the flower with a star. But, if in the same program flow, the user teaches the machine that Figure 43 is a flower, the image will be correctly identified in the next recognition attempt. Through activities like this, the machine learning process can be easily observed and discussed. According to Papert’s Constructionism, the understanding of this process becomes more effective because the system observer was also its builder. One can realize, from concrete to abstract, that the machine learns the way we program it to learn, learns with what we want it to learn (given the paradigm limitations), and does what we allow it to do through the programs we create.

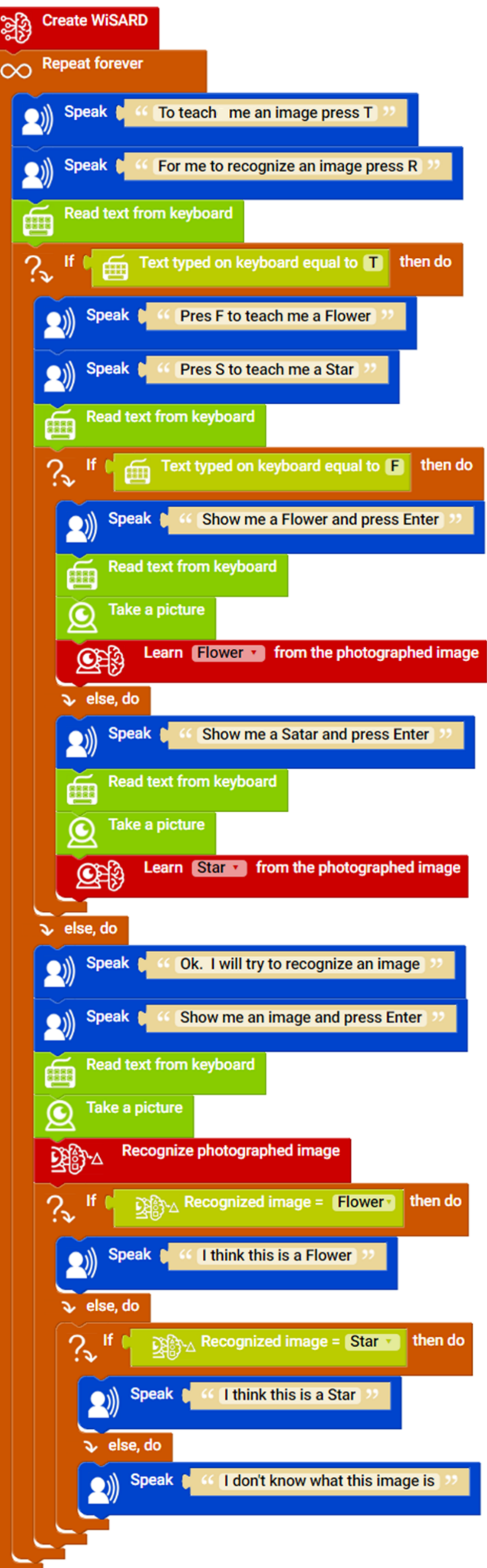

Figure 40: A simple learning machine program built with BlockWiSARD

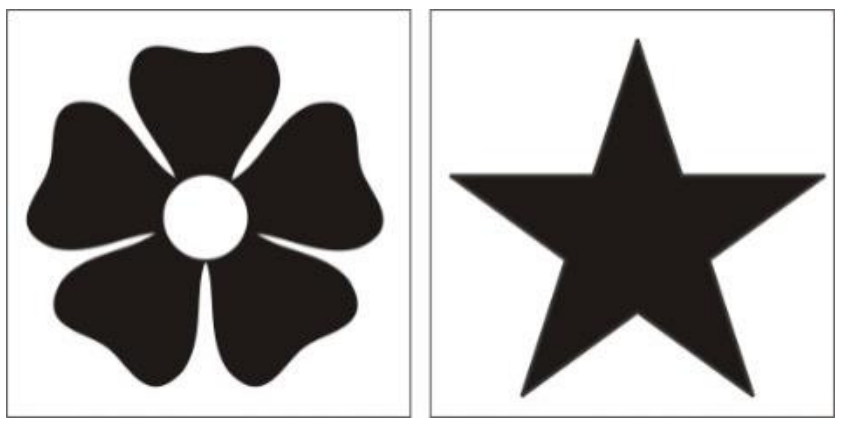

Figure 41: Images used to teach the computer

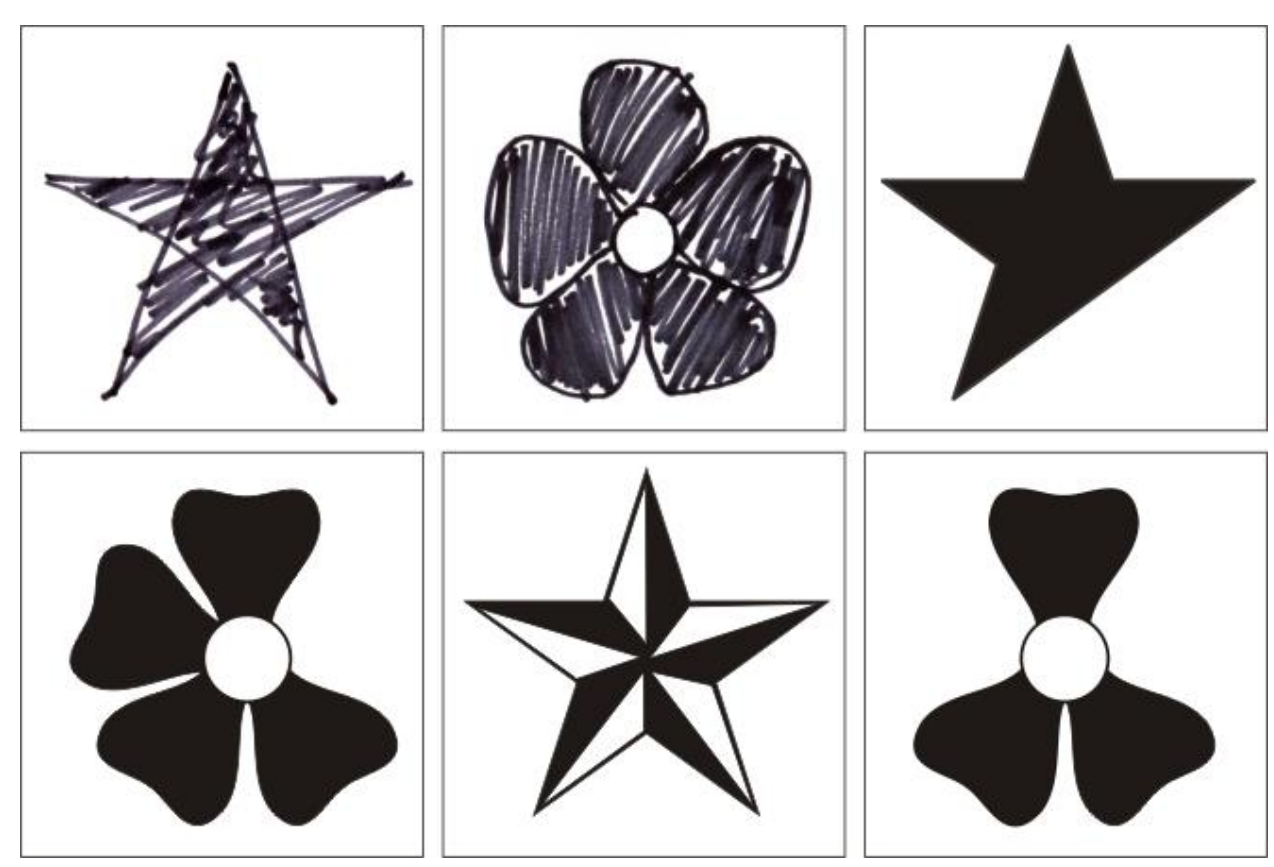

Figure 42: Correctly classified images

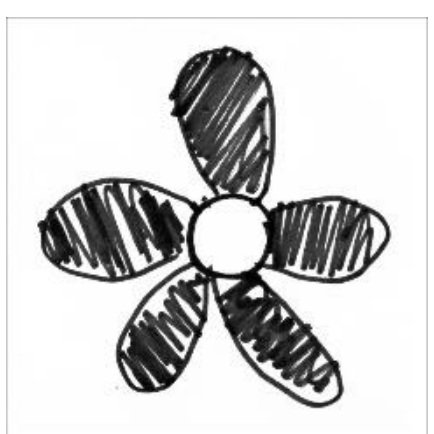

Figure 43: Misidentified image

By connecting an Arduino board to the computer, people can develop systems that controls actuators and acquires data from sensors connected to the board. Using a wheeled robot equipped with a Raspberry Pi[15], a webcam, a speaker and other sensors and actuators, one can build a smart robotic agent that learns in batch or online, and, for example, moves around the environment guided by signs with pictures (Figure 44).

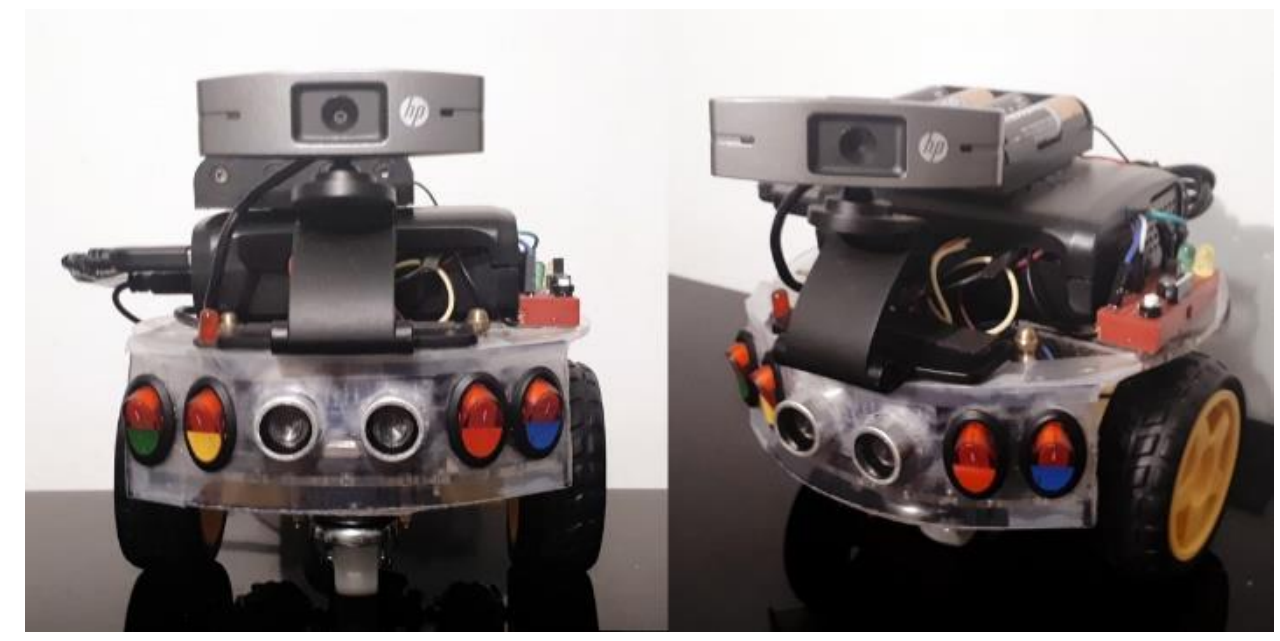

Figure 44: Wheeled robot programmable with BlockWiSARD

By doing those exercises, we expect to rase in the student's minds two relevant questions about artificial intelligence related to two groups of AI definitions presented by Russell and Norvig when discussing the four possible definitions for AI (Figure 45): *Are these machines* *acting rationally? Are they acting like a human?* These are questions that may encourage a debate able to expand the understanding of AI from practice.

**Acting Humanly**

**"The study of how to make computers do things at which, at the moment, people are better."** (Rich and Knight, 1991)

**"The art of creating machines that perform functions that require intelligence when performed by people."** (Kurzweil, 1990)

**Acting Rationally**

**"AI . . . is concerned with intelligent behavior in artifacts."** (Nilsson, 1998)

**"Computational Intelligence is the study of the design of intelligent agents."** (Poole et al., 1998)

Figure 45: The two behavior-related groups from four groups of AI definitions that appear in Russel and Norvig (2010)

## 5.5. Ludic activities with WiSARD model

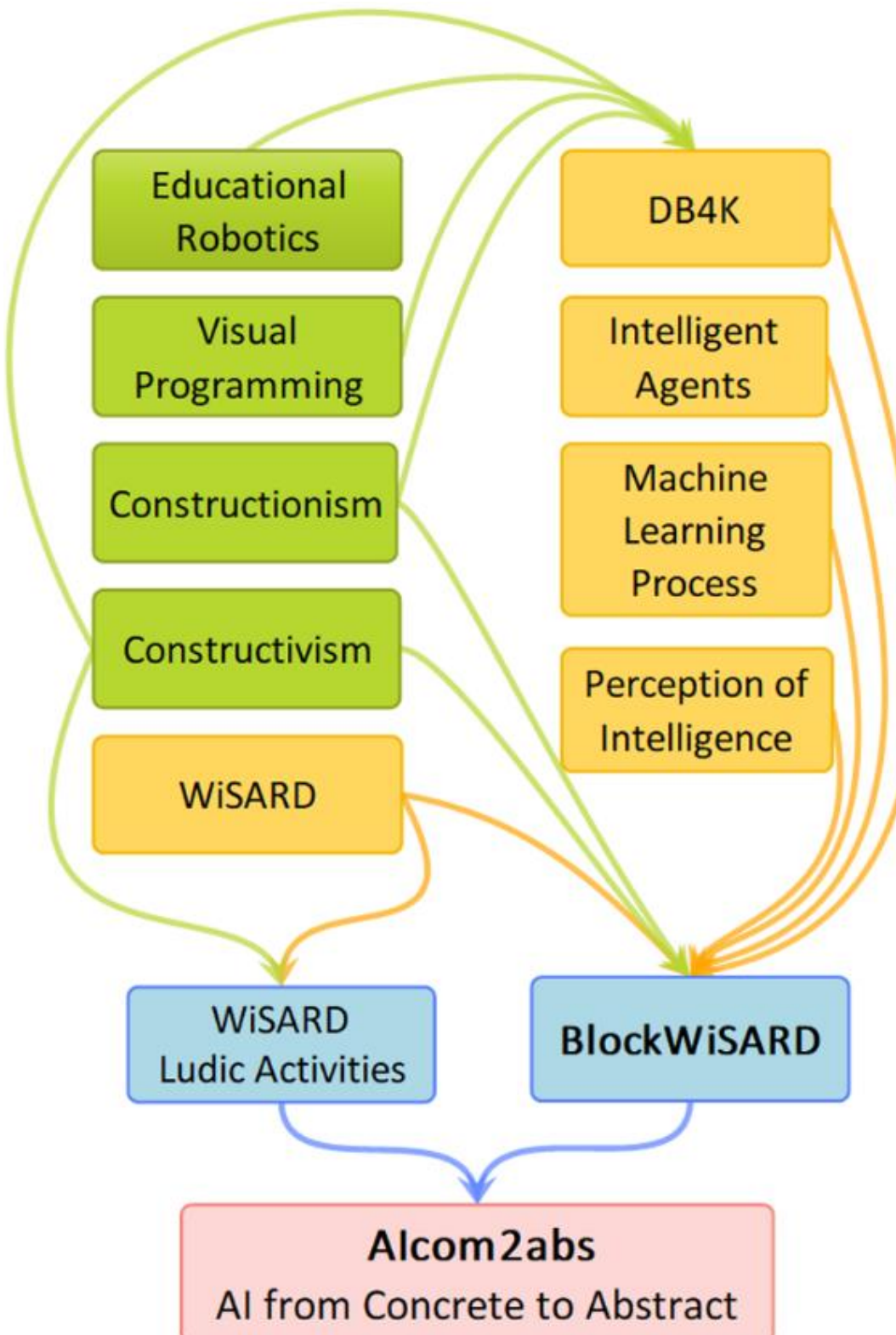

Figure 46: AIcon2abs components diagram

Papert once said:

> The reason you are not a mathematician might well be that you think that math has nothing to do with the body; you have kept your body out of it because it is supposed to be abstract, or perhaps a teacher scolded you for using your fingers to add numbers! This idea is not just metaphysics. It has inspired me to use the computer as a medium to allow children to put their bodies back into their mathematics (Papert 1993b, pp. 31).

WiSARD WANN is an appropriate tool *to put people bodies* into the learning of AI through ludic activities. Along with the use of BlockWiSARD, these activities help to open the *machine learning black* box a little bit more, and advance in the demystification of AI, still starting from the concrete and progressing towards the abstract (Figure 46).

One can view and reproduce, step by step, the training of different classes of images (by writing in the neurons of the discriminators), and later the classification of new image observations (by checking the contents of the neurons of the trained discriminators). A board or a computer game could be advised to make these activities even more fun.

About the concept of consciousness, Turing once said: "[…] the only way by which one could be sure that machine thinks is to be the machine and to feel oneself thinking […]" (Turing 1950, pp. 446). The programs developed with BlockWiSARD help to observe the machine act as if it is thinking. Through the ludic exercises, people can pretend to be WiSARD and check if it is really thinking. From these activities, it is possible to develop a debate about the other two categories of AI definitions presented by Russell and Norvig (Figure 47): *Are these machines thinking rationally? Are these machines thinking humanly*?

**Thinking Humanly**

**"[The automation of] activities that we associate with human thinking, activities such as decision-making, problem solving, learning . . ."** (Bellman, 1978)

**"The exciting new effort to make computers think . . . *machines with minds*, in the full and literal sense."** (Haugeland, 1985)

**Thinking Rationally**

**"The study of the computations that make it possible to perceive, reason, and act."**(Winston, 1992)

**"The study of mental faculties through the use of computational models."** (Charniak and McDermott, 1985)

Figure 47: Two groups related to reasoning and thought processes from four groups of AI definitions that appear in Russel and Norvig (2010)

## 6. Conclusion

The choices that will define the future of artificial intelligence are in the hands of citizens. These choices include the ethical principles on which AI is built and the degree of autonomy to be given to intelligent systems, among others. Thus, it is imperative that people, in general, be aware of what artificial intelligence is. In this way, initiatives that help the general public to build a basic understanding of this domain are of great importance. Inserted in this context, this article presented a methodology that shows the possibility of demystifying AI to the general public by constructing knowledge from concrete references toward abstract concepts.

By including the instantiation of the learning algorithm, and AI learning and classification primitives as components of an otherwise conventional computer program, BlockWiSARD allows the amateur programmer to directly establish the difference between a system that uses artificial intelligence from one that does not. Because WiSARD is able to learn from few examples, it is possible to observe the gain in the performance of the recognition task after the training of each example. Furthermore, AIcon2abs/BlockWiSARD allows the opening of the *machine learning black box* beyond what other approaches do. This is possible because WiSARD has simple learning and classification processes that can be easily observed and replicated. Finally, BlockWiSARD can be used with no internet connection, which is useful for low-income or geographically isolated communities. Table 2 summarizes the features contemplated by AIcon2abs.

Table 2: Summary of AIcon2abs features

| **Features** | **AIcon2abs** |
|---|---|
| Model training is fully performed as part of block-based programming | ✓ |
| Image learning and recognition requires internet | |
| Allows consistent generalizations from a single example | ✓ |
| Online learning process (allows for interleaving between training and classification tasks) | ✓ |
| Uses a third-party AI platform for image training and classification | |
| Uses a simple learning and recognition model that can be replicated in its original form by students in ludic activities | ✓ |
| The learning model adopted allows the visualization, as a concrete image, of what was learned | ✓ |
| The block environment has a downsized set of blocks | ✓ |
| The Block environment has individual personalized icons for each block | ✓ |
| Capable of controlling Raspberry Pi[15] and Arduino[16] GPIOs[17] ? | ✓ |
| Allows image learning and recognition | ✓ |
| Allows text learning and recognition | |
| Allows sound learning and recognition | |
| Allows controlling sprites on computer screen | |
| Text to speech without internet | ✓ |
| Presents theories about cognitive maturity that support the approach in question | ✓ |

Those attributes allow general people, including children, to establish initial contact with the four most adopted groups of AI definitions according to Russel and Norvig (2010): Act Humanly, Act Rationally, Think Humanly, and Think Rationally. The appropriation of these concepts becomes a fertile ground for the debate about artificial intelligence, raising questions like: *Where AI currently stands? Which directions we want AI to take? What aspects of AI are most important to discuss today? What impacts can AI have on our society?* By looking for answers to these questions, people can internalize the main ideas about AI approaches, and develop their critical view on the uses of AI in our society.

This methodology can be applied to both the general public and a more specialized audience, such as students in computer science and engineering courses. Therefore, we hope that the knowledge presented in this research can contribute to building a future in which the benefits of artificial intelligence for society are maximized, and its risks and possible harm can be mitigated as much as possible.

## 6.1. Future work

Physical appearance, expressions, and movements help to define personality traits of a robot. These characteristics significantly influence the way people interact with different robots (Breazeal 2004). Thus, to increase people engagement in the methodology presented here, a study of plastic and behavioral characteristics to be adopted when designing robots used in these processes would be of great value. Besides, for the results of such a study to reach a wider public, it is relevant to consider material costs. So, it is important to identify aesthetic and behavioral aspects that can be incorporated in robots built from the use of cheap or recyclable materials, and low-cost robotics.

Another relevant aspect to be developed from this work is the addition to BlockWiSARD of other WiSARD potentialities such as text, speech, and facial expressions recognition. That may be an interesting way to show people that the same machine learning model can be applied in various domains.

## Acknowledgements

This study was financed in part by the Coordenação de Aperfeiçoamento de Pessoal de Nível Superior - Brasil (CAPES) - Finance Code 001.